\def\gtorder{\mathrel{\raise.3ex\hbox{$>$}\mkern-14mu
    \lower0.6ex\hbox{$\sim$}}}
\def\ltorder{\mathrel{\raise.3ex\hbox{$<$}\mkern-14mu
    \lower0.6ex\hbox{$\sim$}}}

\documentclass[fleqn,usenatbib]{mnras}

\usepackage{newtxtext,newtxmath}

\usepackage[T1]{fontenc}

\DeclareRobustCommand{\VAN}[3]{#2}
\let\VANthebibliography\thebibliography
\def\thebibliography{\DeclareRobustCommand{\VAN}[3]{##3}\VANthebibliography}

\usepackage{graphicx}	
\usepackage{amsmath}	
\usepackage{soul}
\usepackage{caption}
\usepackage{subcaption}

\title[Dissecting Cosmological Filaments at High Redshifts]
  {Dissecting Cosmological Filaments at High Redshifts:\\ 
  Emergence of Spaghetti-type Flow Inside DM Haloes}

\author[Bi, Shlosman and Romano-Diaz]{
Da Bi,$^{1,2}$\thanks{E-mail: dbi224@g.uky.edu}
Isaac Shlosman,$^{2,3,4}$
Emilio Romano-Díaz$^{5}$
\\
$^{1}$Departamento de Astronom\'ia, Universidad de Concepci\'on, Concepci\'on, Chile.\\
$^{2}$Department of Physics and Astronomy, University of Kentucky, Lexington, KY 40506-0055, USA\\
$^{3}$Theoretical Astrophysics, Graduate School of Science, Osaka University, Osaka, Japan\\
$^{4}$Kavli Institute for Theoretical Physics, UC Santa Barbara, CA 93106 \\
$^{5}$Argelander-Institut f\"ur Astronomie, University of Bonn, Auf dem H\"ugel 71, D-53121 Bonn, Germany
}

\date{Accepted XXX. Received YYY; in original form ZZZ}

\pubyear{2022}

\begin{document}
\label{firstpage}
\pagerange{\pageref{firstpage}--\pageref{lastpage}}
\maketitle

\begin{abstract}
We use high-resolution zoom-in simulations to study the fueling of central galaxies by filamentary and diffuse accretion at redshifts, $z \gtorder 2$. The parent haloes were chosen with similar total masses, log\,($M_{\rm vir}/{\textrm M_\odot})\sim 11.75\pm 0.05$, at $z=6$, 4, and 2, in high/low overdensity environments. We analyze the kinematic and thermodynamic properties of circumgalactic medium (CGM) within few virial radii, $R_{\textrm {vir}}$, and down to the central galaxy. Using a hybrid d-web/entropy method we mapped the gaseous filaments, and separated inflows from outflows. We find that (1) The CGM is multiphase and not in thermodynamic or dynamic equilibrium; (2)  filamentary and diffuse accretion rates and densities decrease with lower redshifts, and inflow velocities  decrease from $200-300\,{\textrm {km\,s}^{-1}}$ by a factor of 2; (3) temperature within the filaments increases inside $R_{\textrm {vir}}$, faster at lower redshifts; (4) filaments show a complex structure along their spines: a core radial flow surrounded by a lower density envelope. The cores exhibit elevated densities and lower temperature, with no obvious metallicity gradient in the cross sections. Filaments also tend to separate into different infall velocity regions and split density cores, thus producing a spaghetti-type flow; (6) inside the inner $\sim 30\,h^{-1}$\,kpc, filaments develop the Kelvin-Helmholtz instability which ablates and dissolves them, and triggers turbulence along the filaments, clearly delineating their spines; (7) finally, the galactic outflows affect mostly the inner $\sim 0.5R_{\rm vir}\sim 100h^{-1}$kpc of the CGM.
\end{abstract}




\section{Introduction}
\label{sec:intro}

Galaxies reside inside dark matter (DM) haloes which extend to about ten times the galaxy size. The DM haloes lie within the cosmological network formed by filaments, walls and their intersections, which, together with voids, form the so-called large-scale structure of the universe extensively studied over the last 3--4 decades \citep[e.g.,][]{bardeen86,geller89,bond96,spri06}. Deep within these parent haloes, galactic morphology in contemporary universe has been analyzed for the last century with a great success, solidified by \citet[][see reviews by \citet{binn08,mo10,kormendy13,shlo13}]{hubble36}. 

However, the gray zone of contemporary cosmology lies between these two extremes --- how does the large-scale structure --- the cosmic web, connect to the galactic structure? How do properties of the infalling material change within the halo, and to what extent it preserves its identity on the way to the central galaxy? How does it feed the galaxy growth and affect their morphology. And finally, how does this process evolve with the cosmological time?

In this work, we attempt to answer more modest questions. We focus on comparing the evolution of baryonic filaments and  diffuse accretion in the immediate vicinity of galaxies and their host DM haloes, focusing on the kinematic and thermodynamic properties of the gas. To distinguish the baryonic flows from the extensions of DM filaments inside the haloes, we denote them as {\it streamers}. For this purpose we use our high-resolution cosmological zoom-in simulations presented in \cite{bi22a,bi22b}.  

Two primary modes of galaxy growth exist: mergers with other galaxies and smooth accretion of matter \citep[e.g.,][]{rees77,fall80}. High-redshift galaxies exhibit very high star formation rates for extended periods of time which cannot be supported by galaxy merger events only \citep[e.g.,][]{cham04,keres05,genz06,dekel06,bi22a}. Smooth gas accretion plays an important and even dominant role in the growth of these galaxies, affecting their morphology. In the past decade or so, numerical modeling confirmed that gas accretion can dominate over galaxy mergers at high redshifts, becoming an important component of the hierarchical scenario of galaxy growth \citep[e.g.,][]{keres05,dekel09,devri10,roma11,roma14}.

It is widely accepted that the two complementary mechanisms for gas accretion can operate: the so-called cold and hot accretion modes \citep[e.g.,][]{keres05,dekel06}. \citet{dekel06} have argued that in the hot accretion, the incoming gas, which is supersonic and not virialized, will trigger a shock positioned around the halo virial radius, $R_{\textrm {\textrm vir}}$.  In more recent numerical simulations, the position of this virial shock has been found to lie outside the virial radius \citep[e.g.,][]{nelson16}, and we discuss this point in section\,\ref{sec:discussion4}.  The postshocked gas, which is heated up to the virial temperature, $T_{\textrm {vir}}$, forms a quasistatic envelope, starts to cool and is accreted radially thereafter as it can potentially cool down over less than the Hubble time and hence, contribute as well to the galaxy growth \citep[e.g.,][]{birnb03}. On the other hand, the cold gas penetrates deep inside the halo and streams towards the central galaxy, without being shocked and heated up, allowing for an efficient delivery of a potentially starforming fuel to the central galaxy disk \citep[]{brooks09}. This can even happen when the hot shock is present \citep[e.g.,][]{dekel06,dekel09,Agertz09}. Actually, recent theoretical work has shown that the cold baryonic accretion rather than the hot accretion rate dominates below a certain critical DM halo mass \citep[e.g.,][]{birnb03,keres05,dekel06,Ocvirk08,dekel09,keres09} and the cold accretion rate decays steeply down at lower redshifts \citep[e.g.,][]{keres05,dekel06,roma17}. Therefore, it is especially important to investigate the details of the cold accretion flows in the early universe.  

Several numerical works focused on the cold flows and the baryon cycle of accretion versus outflow in galaxies. Some of them favored low numerical resolution in order to follow an statistical approach to galaxy growth from gas accretion \citep[e.g.,][]{keres05,Ocvirk08,dekel09}, while others used "zoom-in" simulations to address the process on galactic scales at specific redshifts \citep[e.g.,][]{Faucher11,Fumagalli11,Kimm11,Stewart11,Goerdt12,Shen13,roma17}. 

In this work, we focus on the baryonic accretion onto the central galaxy within its parent DM halo at the Cosmic Dawn, ending at three selected redshifts, $z_{\textrm f} = 6$, 4, and 2. Using high-resolution zoom-in simulations \citep{bi22a}, we follow the galaxy evolution within similar mass DM haloes, log\,$M_{\textrm {vir}}\sim 11.75\pm 0.05\,{\textrm {M}_\odot}$, which are only a factor of 2 below those haloes which are the most efficient in producing stars \citep[e.g.,][]{behr13}.   

The chosen mass range ensures that the selected halos can create favorable conditions for both the hot and cold accretion flows, and are expected to form sufficiently massive galaxies to be properly resolved numerically in our simulations. In addition, these halos are expected to contain $L*$ galaxies at their respective $z_{\rm f}$. Hence, they can be compared with present day $L*$ galaxies. Our final redshifts, $z_{\textrm f}$, encompass the end of the reionization epoch, $z\gtorder 6$, and the subsequent time period of $\sim 2.5$\,Gyr, when the star formation in the universe peaks. Overall, galaxy evolution for $z\sim 9-2$ is analyzed.

Our main goal is to study accretion of the cosmological gas from the large scale, i.e., from $\sim 4R_{\rm vir}$, down to the central galaxies, where the gas settles in the circumgalactic space. We define this gas as the circumgalactic medium \cite[CGM, e.g.,][for a recent review]{tomlin17}. This basically corresponds to the gaseous component of the DM halos \citep[e.g.,][]{sadoun19},  but its radial extension will be more precisely defined in sections\,\ref{sec:results4} and \ref{sec:discussion4}. and we analyze its specific properties, i.e., the kinematics, chemical composition and thermodynamics. Within this spatial range, the gas transform from the extragalactic medium to the starforming ISM. This gas is expected to supply the new material for star formation from the IGM, satellite galaxies, the hot and cold accretion phases, and from the recycled material injected by the disk stars and active galactic nuclei (AGN) --- as a result of the galactic feedback processes. This region also has the potential of hiding the missing baryons in the universe. Moreover, the CGM dynamic and thermodynamic states can be altered by the stellar and AGN feedback mechanisms which in this case will be responsible for the quenching of star formation in galaxies \citep[e.g.,][]{davies20}.

Major efforts have been aimed at understanding the feedback of winds from massive stars, supernovae (SN), and AGN on galaxy evolution. However, specific details of this feedback and its implementation and fine tuning to observational and computational models are far from being understood \citep[e.g.,][]{yajima15,sadoun16}.  In particular, the spatial extent of this feedback is still unclear --- is it limited to the inner DM haloes, to the halo virial radius, or extends beyond it? 

This paper has been organized as follows. Section\,\ref{sec:numerics4} describes the numerical issues and the methods used. Section\,\ref{sec:results4} presents our results. This is followed by the discussion section and the summary.

\section{Numerical Modeling}
\label{sec:numerics4}
\subsection{Simulation Suite}
\label{sec:sims4}

We analyze the zoom-in simulations suite presented in \citet{bi22a}  which used the hybrid $N$-body/hydro code \textsc{gizmo} \citep{hopk17} with the Lagrangian meshless finite mass (MFM) hydro solver. The full details of the simulations are given in \citet{bi22a}. Here we only provide the highlights. 

The simulations adopted the \cite{planck16} $\Lambda$CDM concordant model, i.e. $\Omega_{\textrm m} = 0.308$, $\Omega_\Lambda = 0.692$, $\Omega_{\textrm b} = 0.048$, $\sigma_8 = 0.82$, and $n_{\textrm s} = 0.97$. 
The Hubble constant is taken as $h = 0.678$ in units of $100\,{\textrm {km}\,\textrm s^{-1}\,\textrm {Mpc}^{-1}}$. 

Uni-grid, DM-only initial conditions (IC) were generated at redshift $z=99$ by means of the \textsc{music} code \citep{hahn11} within a box of $50 h^{-1}$\,Mpc, which were evolved until redshift 2. From this parent simulation haloes of the same mass range were chosen at three different target redshifts, $z_{\textrm f} = 6$, 4, and 2, to be re-simulated at much higher resolution. The selected DM haloes have been chosen to reside at low and high overdensities, $\updelta\sim 1$ and $\updelta\sim 3$, respectively \citep{bi22a}. 

The individual zoom-in ICs, also generated with \textsc{music}, were composed of five nested levels of refinement on top of the base grid, i.e., from $2^7$ to $2^{12}$. Their DM-only versions were first evolved in order to check for (and to avoid) contamination from massive, lower-resolution particles at the highest resolution-level volume. After this, baryons were included at the highest level of refinement in the reconstruction of their respective ICs. All the details of the models are displayed in the Table\,\ref{tab:DMsim}, as well as in Tables\,1 and 2 of \citet{bi22a}. The total number of selected haloes were 6, and in combination with the two different galactic winds feedbacks (see section\,\ref{sec:winds4}) the final simulation suite was composed of 12 haloes.

Within this setup, the effective number of particles (DM and baryons) in our simulations is $2\times 4096^3$, leading to a mass resolution per particle of $3.5\times 10^4\,{\rm M_\odot}$ for gas and (eventually) stars, and $2.3\times 10^5\,{\rm M_\odot}$ for DM. The minimal adaptive gravitational softening (in comoving units) for gas was 74\,pc, and for stars and DM 74\,pc and 118\,pc. This means that at the final redshifts, $z_{\rm f}=6$, 4, and 2, the softening for stars in physical coordinates is 10.5\,pc, 14.7\,pc, and 24.6\,pc, respectively. 

For better angular momentum conservation and in order to resolve the Kelvin-Helmholtz instability which is expected to develop when the cosmological filaments penetrate the DM halo, the MFM hydro solver was employed instead of the “traditional” SPH solver with an adaptive gravitational softening for the gas. The hybrid multiphase model was invoked for the ISM and star formation \citep{spri03}. In this, starforming particles contain the cold phase that forms stars, and the hot phase that results from the SN\,II heating. Metal enrichment is included: the metallicity increase in the starforming gas scales with the fraction of gas in the cold phase, the fraction of stars that turns into SN, and the metal yield per SN. A total of 11 metal species were followed in both gas and stars, including H, He, C, N, O, Ne, Mg, Si, S, Ca, and Fe. Metal diffusion is not implemented explicitly, but metals can be transported by winds (see section\,\ref{sec:winds4}). The density threshold for star formation (SF) was set to $n^{\textrm {SF}}_{\textrm {crit}} = 4\,{\textrm {cm}^{-3}}$.  Furthermore, our simulations include the redshift-dependent cosmic UV background, which has been turned on at $z = 11.7$ \citep{Faucher09}.

\subsection{Identification and Properties of DM haloes and Galaxies }
\label{sec:group}

DM haloes and their properties were identified by the group finder \textsc{rockstar} \citep{behr13}, with a Friends-of-Friends 
linking length of $b=0.28$. The halo virial radius and the virial mass, $R_{\textrm {vir}}$ and $M_{\textrm {vir}}$, have been defined by $R_{200}$ and $M_{200}$ \citep[e.g.,][]{nfw96}. $R_{200}$ is the radius within which the mean interior density is 200 times the critical density of the universe at that time. 
 
Galaxies have been identified by the group-finding algorithm \textsc{hop} \citep{eise98}, using the outer boundary threshold of baryonic density of $10^{-2}\,n^{\textrm {SF}}_{\textrm {crit}}$, which ensured that both the host starforming gas and the lower density non-starforming gas are roughly bound to the galaxy \citep{roma14}. This assures that identified galaxies are not imposed with a particular geometry. Note that with this definition, all the galaxies appear to be generally smaller than galaxies defined with $0.1R_{\textrm {vir}}$, which is typically used in the literature \citep[e.g.,][]{scanna12,mari14}.

\subsection{Galactic Wind Models and Supernovae Feedback}
\label{sec:winds4}

We also have made used of the two wind models of \citet{bi22a} --- the Constant Wind \citep{spri03} and the Variable Wind \citep{oppe06} (hereafter CW and VW, respectively). Both wind models implement the concept of "the decoupled particle wind," when a wind particle decouples from the ambient gas particles, no longer interacts hydrodynamically, and move ballistically.  The time period of decoupling stage depends on the shortest of either $10^6$\,yrs, or when the particle moves to a region (but still within the galaxy) where the background gas density is lower by a factor of 10 compared to the critical density  of star formation. 

Both CW and VW orientations have been assumed isotropic. For the CW model, the wind velocity is $v_{\textrm w} = 484\,{\textrm {km}\,\textrm s^{-1}}$, and the mass loading factor, $\beta_{\textrm w}\equiv \dot M_{\textrm w}/\dot M_{\textrm {SF}} = 2$ \citep{sadoun16}. Here $\dot M_{\textrm w}$ is the mass loss rate by the wind, and $\dot M_{\textrm {SF}}$ is the star formation rate (SFR) measured in $M_\odot\,{\rm yr^{-1}}$. For the  VW model, the wind velocity scales with the physical escape velocity of the host halo. In this case, the mass loading factor has been calculated assuming the total wind energy, and it is given by the energy-driven and the momentum-driven winds. In general, by calculating the total kinetic energy of all the wind particles, the VW constitutes a stronger feedback, about 8 times stronger on the average than the CW. Yet, at each moment of evolution this ratio can fluctuate in both directions.

In addition to galactic winds, this modeling includes the SN feedback \citep{spri03}. The SN deposition of energy in the gas is fixed at $4\times 10^{48}\,{\rm erg\,M_{\odot}^{-1}}$. The rate of the SN depends on the initial stellar mass function (IMF), which is taken as the Chabrier IMF \citep{chabrier03}.

\subsection{Cosmic Web Decomposition}
\label{sec:web}

The cosmic web can be divided into its various components, i.e., voids, sheets/walls, filaments, and clusters/knots \citep[e.g.,][]{Lapparent86,colless03,tegmark04,mehmet14}. A number of different methods have been attempted in the literature in both numerical simulations and observational data to separate the web into components \citep[]{argon07a,hahn07a,forero09,bond10,sousbie11,hoffman12,cautun13}. We have employed a hybrid scheme, involving the d-web, which is based on the orbital stability using the local tidal tensor usually calculated for the DM distribution \citep[]{hahn07b}, as well as two additional methods discussed below. Because we are mainly interested in the evolution of the baryonic component within DM haloes, this method has been {applied to scales smaller  than few$\times R_{\rm vir}$ and to the baryonic distribution, rather than to the DM one.   In this way, we have been capable to follow the baryonic streamers, i.e., cold streams, inside the virial radius \citep[e.g.,][]{dekel09}. 

We briefly describe the d-web method. A particle {\it i} moving in a peculiar gravitational potential, $\phi_i$, can be described by the following equation of motion in comoving coordinates,
\begin{equation}
\ddot{x}_\textrm i=-\bigtriangledown  \phi_\textrm i \,,
\end{equation}
where the dots represent derivatives with respect to time. After linearizing the equation of motion, the system can be re-written as
\begin{equation}
\ddot{x}_\textrm i=-T_{\textrm {ij}}(\bar{x}_\textrm k)(x_\textrm j-\bar{x}_{\textrm {k,j}}) \,,
\end{equation}
where ${T_{\textrm {ij}}}$ represents (in our case) the tidal tensor of baryonic gravitational potential within the DM haloes. By calculating the Hessian\footnote{The Hessian matrix or Hessian is a square matrix of second-order partial derivatives of a scalar-valued function, or scalar field.} matrix, we can define different matrix elements by an analogy with the Zel’dovich approximation \citep{zel'70}, which depend on the basis of the eigenvalues $\uplambda_1$, $\uplambda_2$, $\uplambda_3$ of the tidal tensor,

\begin{itemize}
\item voids: $\uplambda_1$ < $\uplambda_2$ < $\uplambda_3$ < 0
\item sheets/walls: $\uplambda_1$ < $\uplambda_2$ < 0 < $\uplambda_3$
\item filaments: $\uplambda_1$ < 0 < $\uplambda_2$ < $\uplambda_3$
\item knots: 0 < $\uplambda_1$ < $\uplambda_2$ < $\uplambda_3$
\end{itemize}

However, the particular details of this classification method, i.e., extension, thickness of the identified structures, depends on the length scale of the potential field's grid resolution.  

As our goal is to extend the filament identification to the baryonic component inside the virial radius, we complement the d-web method outside the virial radius with an entropy-based method inside the such radius, as described below.

We define the entropy as $K=T/\rho^{\frac{2}{3}}$, which is computed using the effective equation of state. The virial entropy of the halo, $K_{\textrm {vir}}$, is defined by the virial temperature and the total average density at the virial radius  \citep[e.g.,][]{dekel09}. In this way, the penetrating cold (filamentary) baryon streams on this scale show a lower entropy than the virial entropy of the halo, $K_{\textrm {vir}}$.  

We use the d-web outside the virial radius where the potential has been smoothed by $10h^{-1}$kpc Gaussian filter in order to define the entropy cuttoff there. The entropy cutoff constant has been defined as the maximal entropy which results in similar filaments outside the virial radius obtained from the d-web method. This cutoff is used inside the virial radius as well. We have tested this method, and it shows a smooth transitions across the virial radius.  Furthermore, we extended the baryonic d-web determination inside $R_{\rm vir}$, down to $50h^{-1}$kpc, in tandem with the entropy method, for comparison.  In addition, inside the virial radius, we apply the baryon kinematics method to distinguish the accretion flow from the outflow, in the region where both coexist.

We have tested the above hybrid method and found that it successfully extrapolates the d-web/entropy elements down to $\sim 10h^{-1}$\,kpc, where the boundary of the galaxy lies.  

\section{Results}
\label{sec:results4}

We present our results based on our full simulation suite, starting with the detection of cold baryonic streamers and diffuse accretion. Furthermore, we emphasize differences between the two galactic wind feedback models used in our simulations under identical initial conditions, the CW and VW galactic outflow models, and the environmental effects.

\subsection{Identifying the filamentary streamers}
\label{sec:results41}

By using our hybrid d-web/entropy method (section\,\ref{sec:web}), we have been able to separate the large scale flows (outside  $R_{\rm vir}$) down to the central galaxy.  

Figure\,\ref{fig:allfil} displays the mapped individual streamers  within a box of side $1,200h^{-1}$\,kpc, as computed with the d-web algorithm. Only CW feedback models are shown because at these scales there is no observable difference between both feedback schemes. All models are displayed at their final redshifts.

We follow the filamentary streamers and diffuse accretion flows from  $\sim 3R_{\textrm {vir}}$ down to the inner $\sim 10\,h^{-1}$\,kpc from the halo centers. Our general implementation allow us to overcome the difficulty in following the accretion streams deep into the DM haloes and connecting them to galaxies. Furtheremore, the hybrid d-web/entropy method supplemented by the kinematics inside $R_{\rm vir}$ allows us to separate the inflow from outflow. The streamers have been concatenated to the innermost flow using the baryons kinematic (see Appending\,{sec:append}). We discuss the details of gas motions in the innermost regions and their penetration into the central galaxies in section\,\ref{sec:results44}.

\begin{figure}
\center 
	\includegraphics[width=0.48\textwidth]{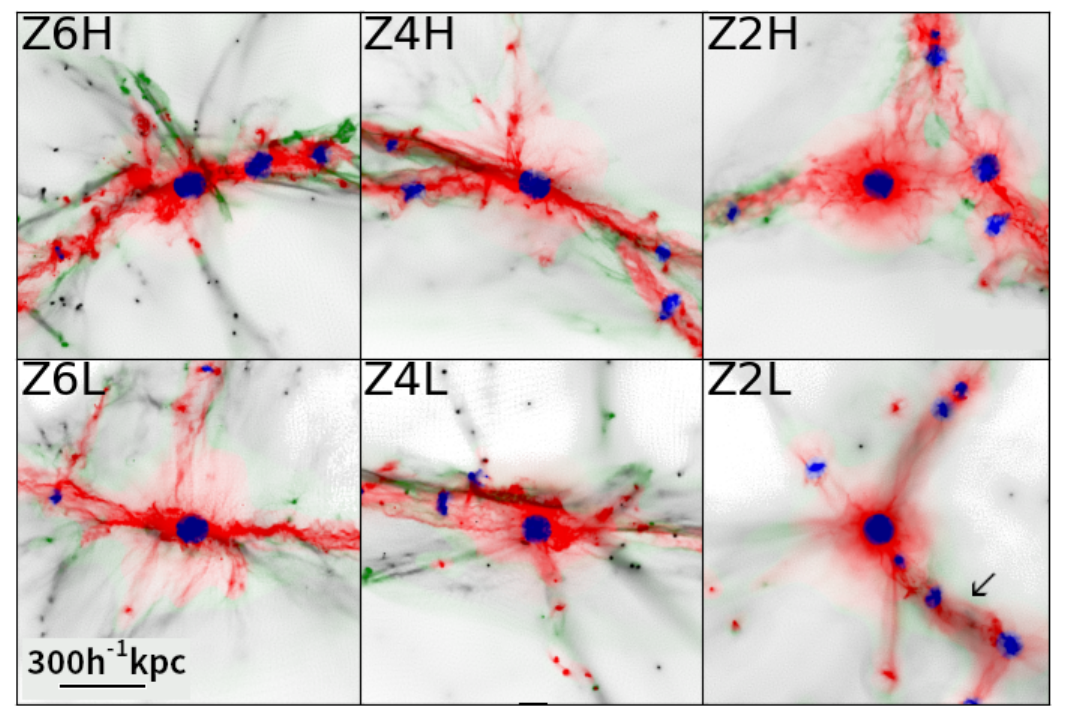}
    \caption{Structure classification of the DM-haloes environment using the   d-web method applied to baryons in boxes of length $1,200\,\textrm h^{-1}$\,kpc, ($\sim \pm 3R_{\rm vir}$).  The color coding represents: knots in blue, filaments, (i.e., streamers) red, walls green and voids in black.  From left to right columns, 
    haloes at their final redshifts $z_{\textrm f} = 6$, 4, and 2 are shown. All the models correspond to the CW feedback as they are similar to the VW models at these scales. The upper panels display haloes in the high overdensity environments and the lower panels show haloes in low overdensity environments, (see Table\,\ref{tab:DMsim}). The arrowhead at the Z2L snapshot points to the filament \#1 which is analyzed in detail in the subsequent sections.}
    \label{fig:allfil}
    \end{figure}

\begin{table}
\resizebox{\columnwidth}{!}{
\vspace*{-.1cm}
\centering
\begin{tabular}{cccccc}
\hline
 $\mathrm{z_{\rm f}}$ & Model Name &  ${\rm log}\,M_{\rm vir}\,$M$_\odot$ & $R_{\rm vir}\,{\rm kpc}$ & $\updelta$  \\
\hline
\hline   
    6     & Z6H & 11.7   & 184$h^{-1}$ &  3.04   \\
\hline   
    6     & Z6L & 11.7  & 184$h^{-1}$ &  1.60   \\
\hline 
\hline   
    4     & Z4H & 11.7  & 184$h^{-1}$ &  3.00   \\
\hline   
    4     & Z4L & 11.8   & 185$h^{-1}$ &  1.33   \\
\hline
\hline   
    2     & Z2H & 11.8   & 206$h^{-1}$ &  2.80   \\   
\hline      
    2     & Z2L & 11.8   & 195$h^{-1}$ &  1.47   \\    
  
\hline   
\hline
\end{tabular}
}
\caption{Table of the halo properties in the DM only simulations. All values are given at the final redshifts $z_{\rm f} = 6$, 4, and 2. The columns correspond to (from left to right): the final redshift $z_{\rm f}$; the model number (see definition in section\,\ref{sec:group}; the virial mass of DM halo $M_{\rm vir}$ at $z_{\rm f}$; the halo virial radius (in comoving coordinates) $R_{\rm vir}$; $\updelta$ --- the local overdensity.  } 
\label{tab:DMsim}
\end{table}

We observe that, overall, the DM haloes are typically associated with two or three main cosmological filaments at $z_{\rm f}$, as shown in Figure\,\ref{fig:allfil}.

\begin{figure*}
\center 
	\includegraphics[width=0.85\textwidth]{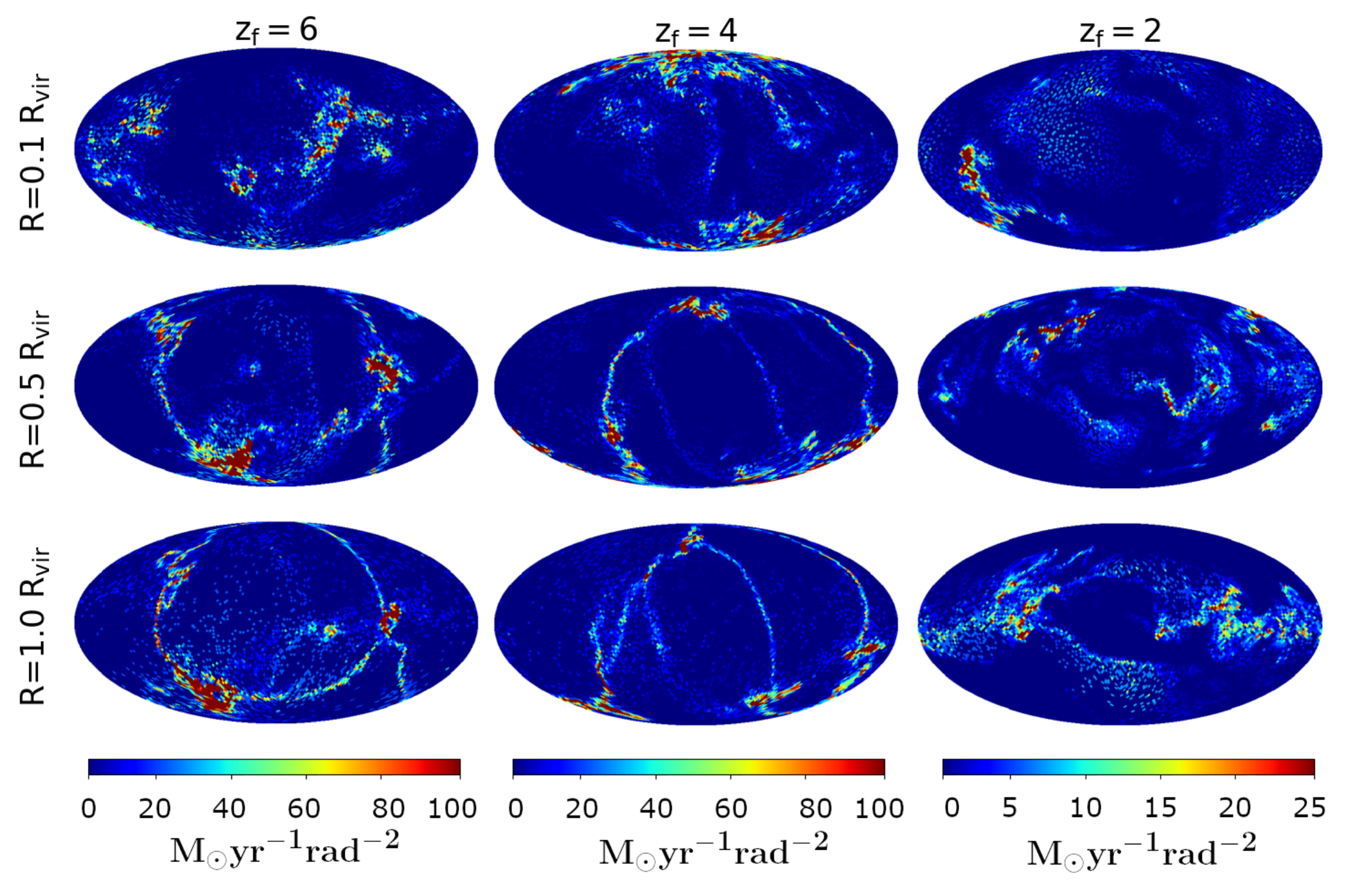}
    \caption{Inflowing streamers, diffuse accretion, and walls (seen as threads connecting the streamers) at three representative radii in a thick shell of $10h^{-1}$\,kpc, at 0.1$R_{\rm vir}$, 0.5$R_{\rm vir}$ and 1$R_{\rm vir}$ for haloes at $z_{\textrm f} = 6$, $z_{\textrm f} = 4$ and $z_{\textrm f} = 2$. Shown as whole-sky HEALPix projection maps (see the text). The color represents radial mass influx of gas per solid angle. Thin walls are seen between the streamers at larger radii until dissolved on the galaxy scale of $\sim 0.1 R_{\rm vir}$, especially at $z_{\textrm f} = 6$ and $z_{\textrm f} = 4$. }
    \label{fig:skymap}
\end{figure*}

\subsection{Radial properties of filamentary streamers}
\label{sec:results421}

\begin{figure*}
\center 
	\includegraphics[width=0.95\textwidth]{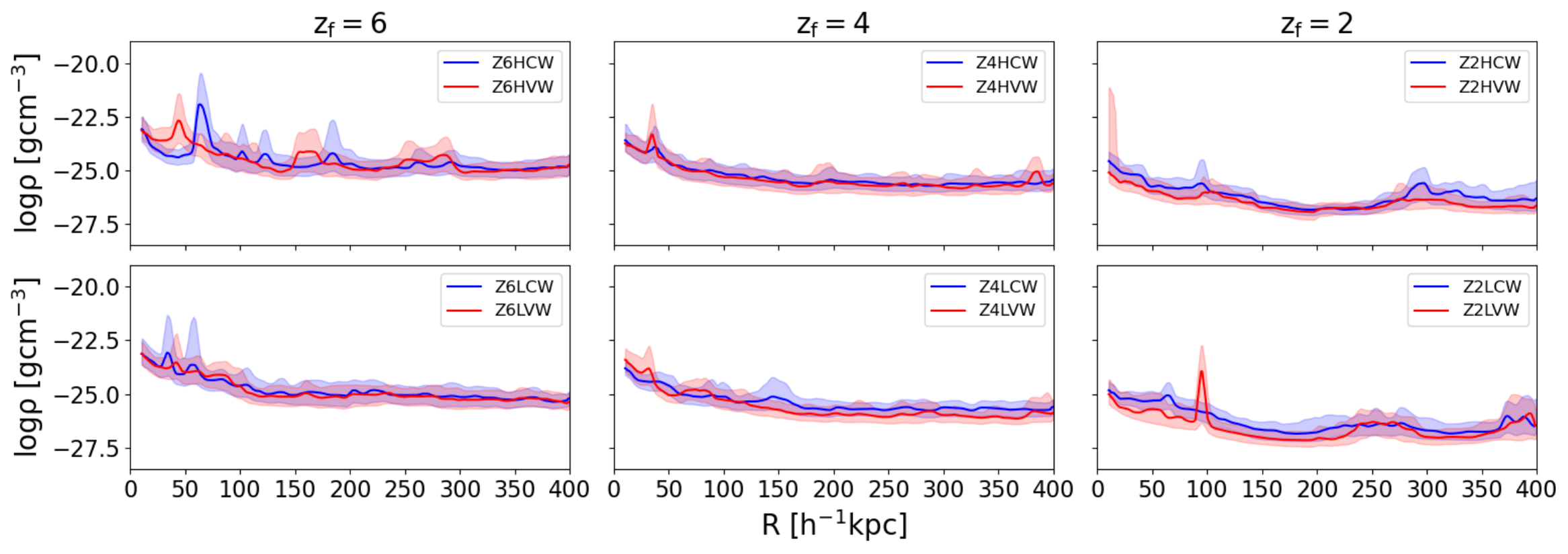}
    \caption{Volume density radial profiles of gas representing the sum of all the streamer spines at $z_{\textrm f} = 6,$ 4, and 2, from left to right columns, respectively. The streamers have been divided into shells of $1\,h^{-1}$\,kpc thickness.  The corresponding DM haloes can be identified from Figure\,\ref{fig:allfil}.  The VW feedback models are given by the red lines and CW models by the blue lines, which represent the median. The color shadows show the 20-80 percentile scale.  The upper panels display haloes in the high overdensity environment and the lower panels show haloes in the low overdensity environment (see Table\,\ref{tab:DMsim} for further details). }
    \label{fig:allfilarho}
    \end{figure*}    

 \begin{figure*}
\center 
	\includegraphics[width=1\textwidth]{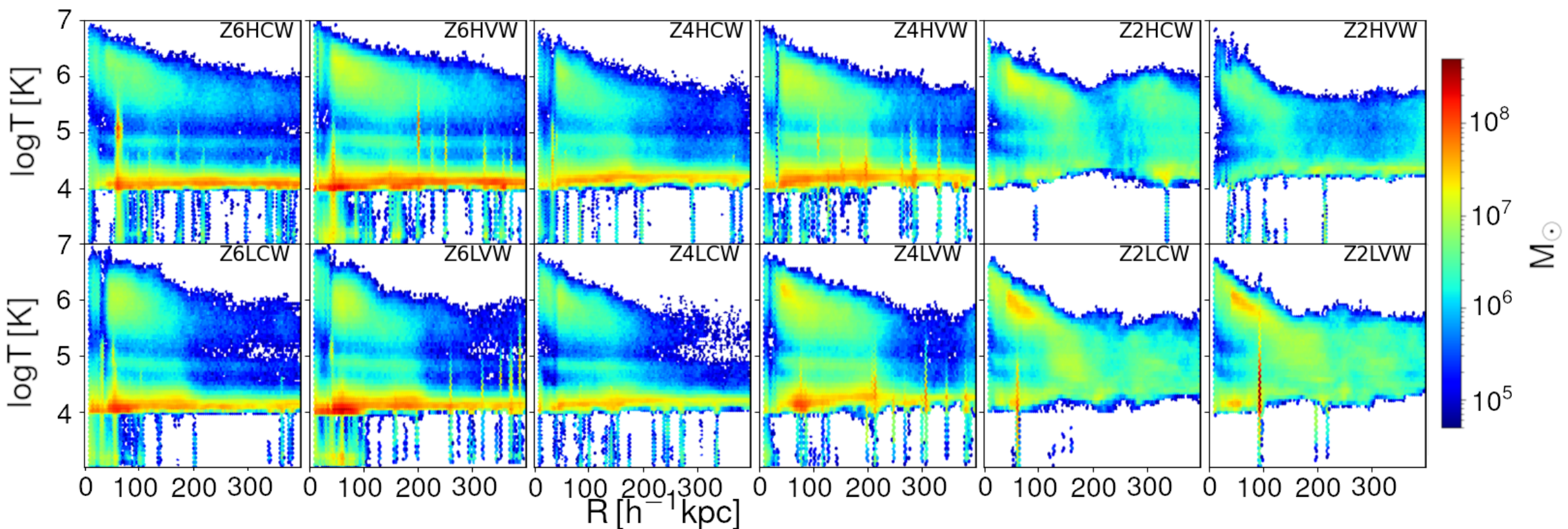}
    \caption{The hex-binned gas temperature radial profiles of the filamentary streamers within $2R_{\rm vir}$ at $z_{\rm f} = 6$, 4 and 2, with  the color palette representing the gas mass in individual pixels.  Note that the virial temperatures of these DM haloes is $T\sim 1.3\times 10^6(1+z)$\,K.}
    \label{fig:allfilaT}
    \end{figure*}    

 \begin{figure*}
\center 
	\includegraphics[width=0.95\textwidth]{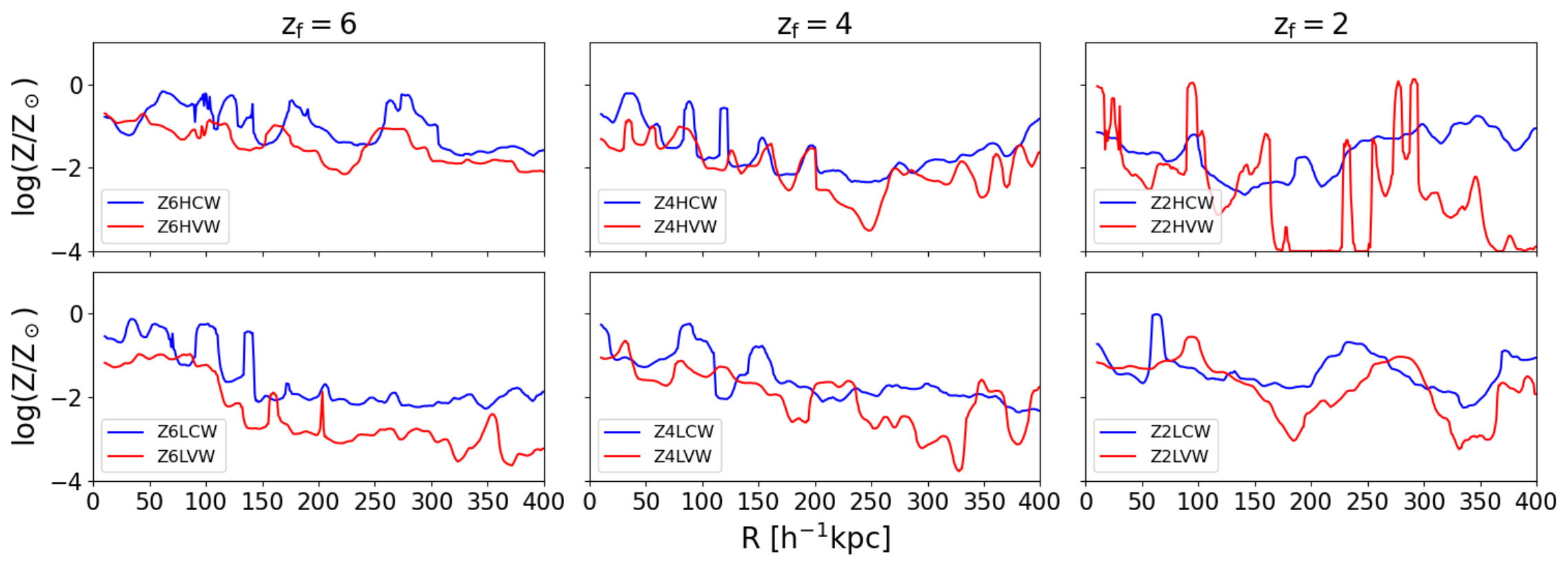}
    \caption{As in Figure\,\ref{fig:allfilarho}, but for the mass weighted average gas metallicity radial profiles of filamentary streamers. }
    \label{fig:allfilaZ}
    \end{figure*}    

 \begin{figure*}
\center 
	\includegraphics[width=0.95\textwidth]{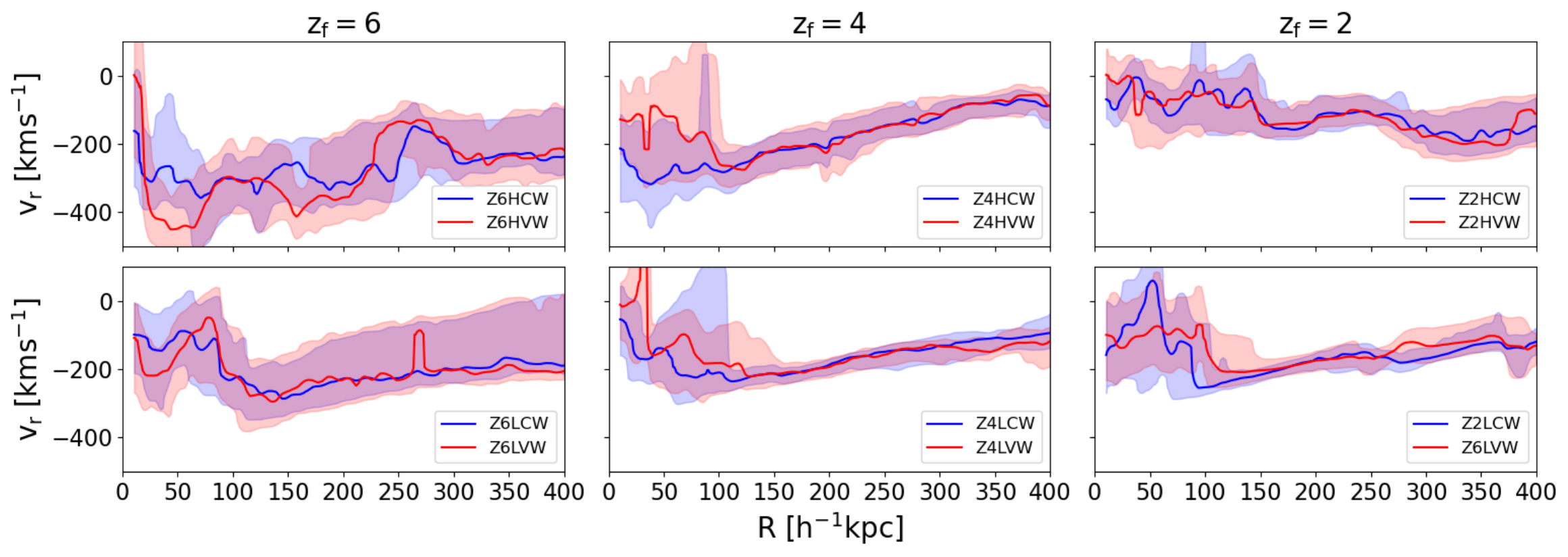}
    \caption{As in Figure\,\ref{fig:allfilarho}, but for the filamentary gas inflow radial velocity profiles. }
    \label{fig:allfilavr}
    \end{figure*}

 \begin{figure*}
\center 
	\includegraphics[width=0.95\textwidth]{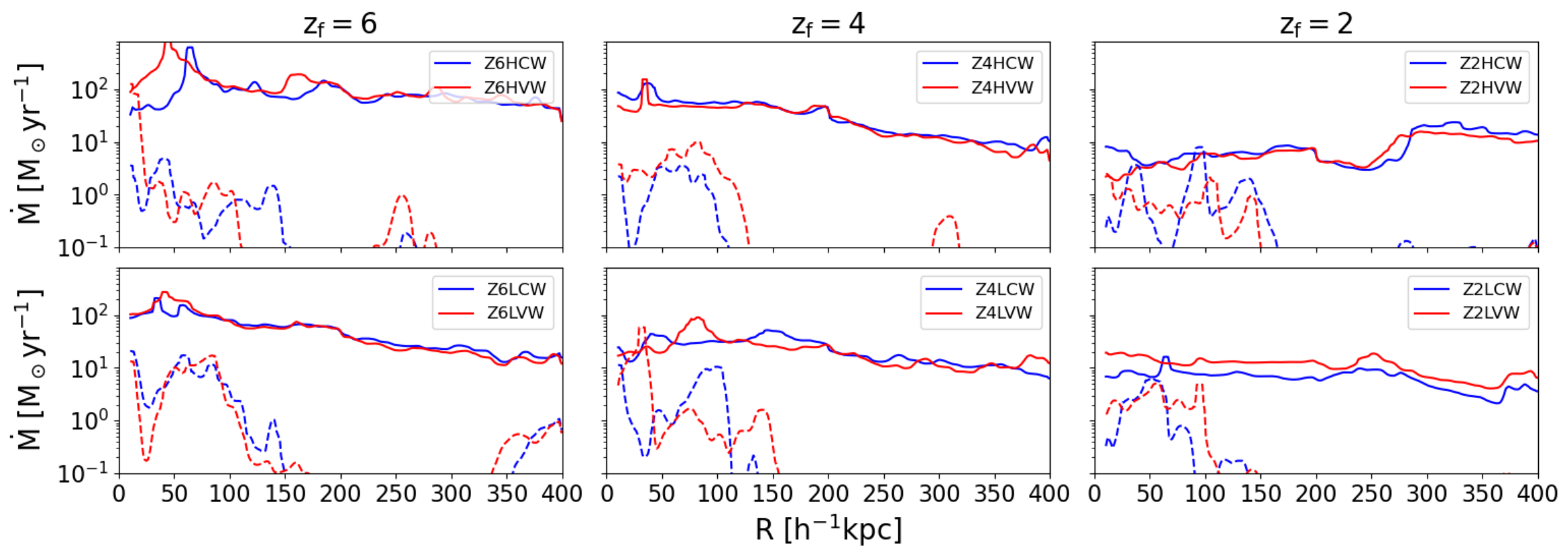}
    \caption{Radial profiles of the mass accretion rates along the filamentary streamers (solid lines) and of the outflows along the same structures (dashed lines).}
    \label{fig:allfilaflux}
    \end{figure*}

 \begin{figure*}
    \centering 
\begin{subfigure}{0.45\textwidth}
  \includegraphics[width=\linewidth]{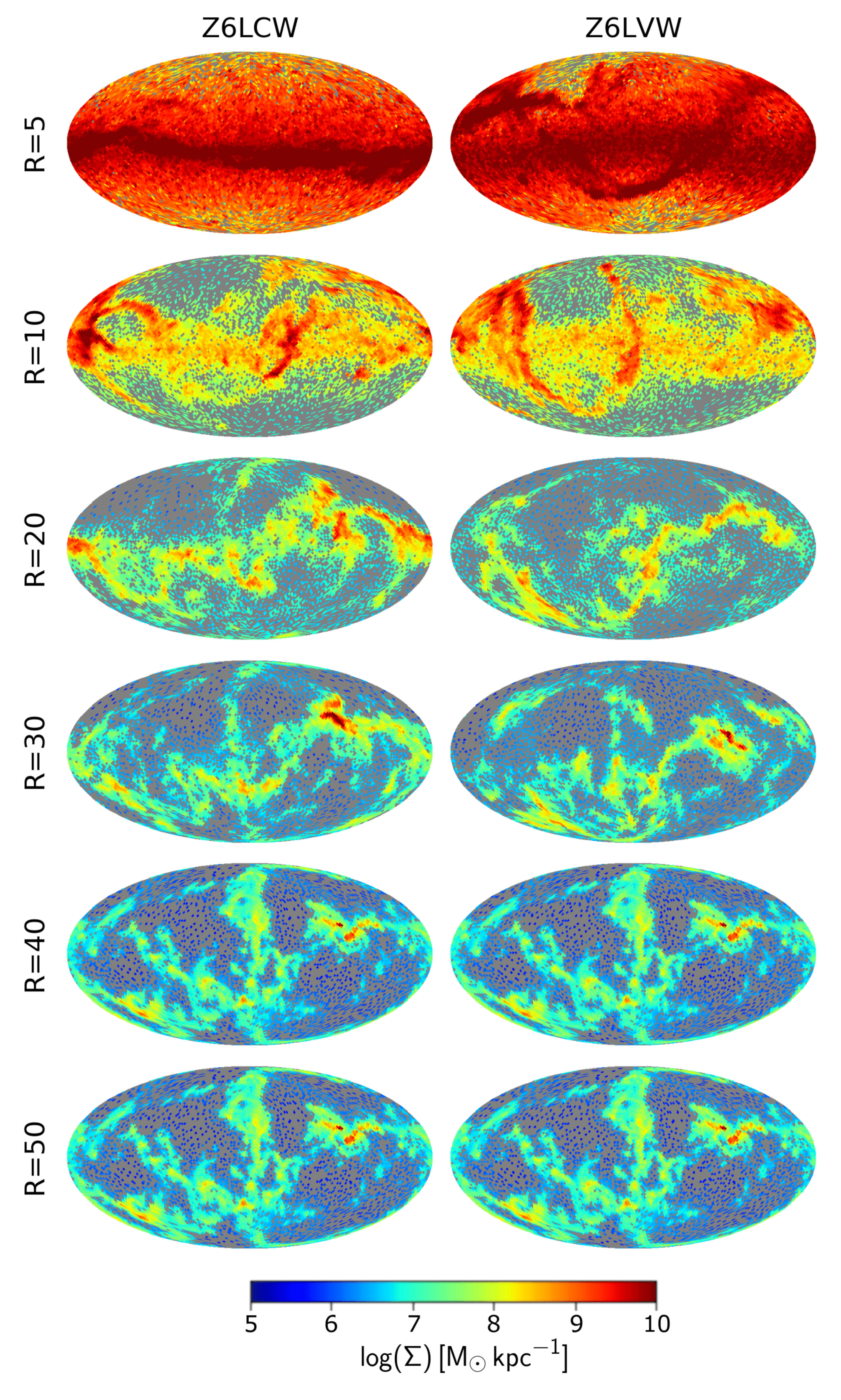}
  \caption{}
  \label{fig:1}
\end{subfigure}\hfil 
\begin{subfigure}{0.43\textwidth}
  \includegraphics[width=\linewidth]{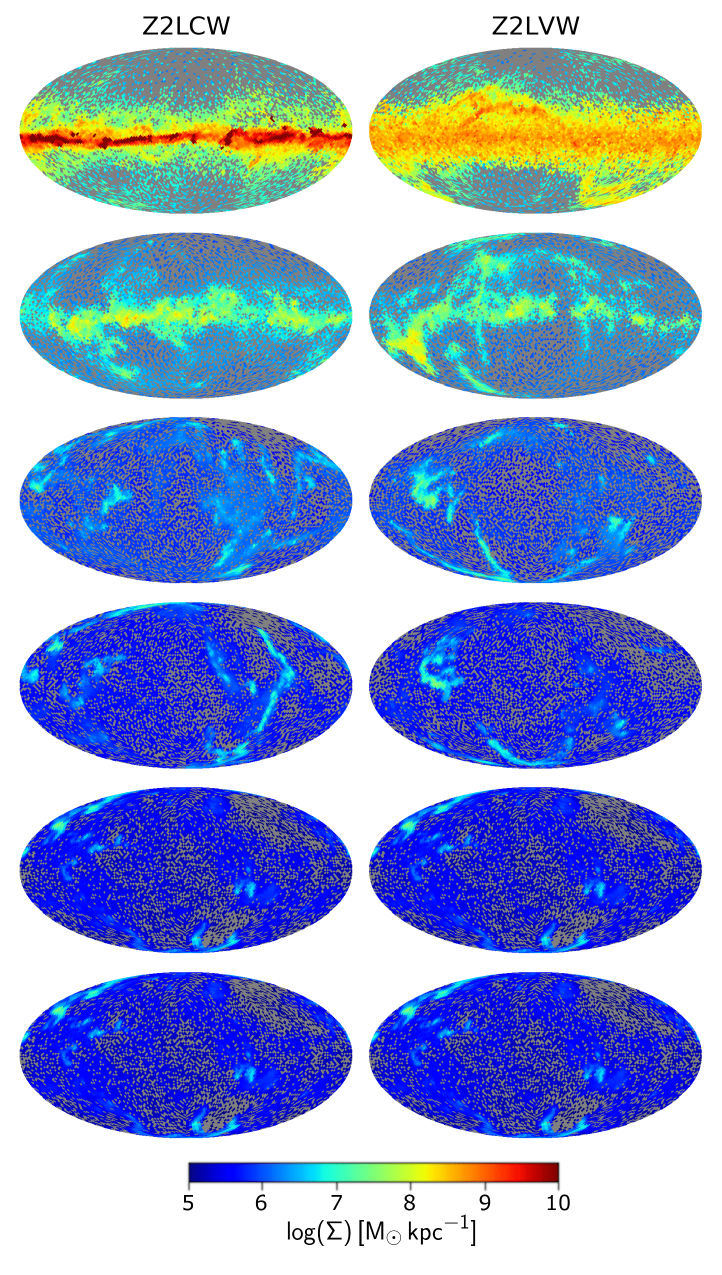}
  \caption{}
  \label{fig:2}
\end{subfigure}\hfil 

\caption{Whole-sky HEALPix projection maps (see text) of the inflowing streamers and diffuse accretion in $3h^{-1}$kpc thick shells at 5, 10, 20, 30, 40 and 50$h^{-1}$kpc for haloes at $z_{\textrm f} = 6$ (a) and $z_{\textrm f} = 2$ (b) with both CW (left column) and VW (right column). The color bar represents the gas surface density. }
\label{fig:surfdens}
\end{figure*}   

Figure\,\ref{fig:skymap} shows the whole-sky Hierarchical Equal Area isoLatitude Pixelisation (HEALPIX) projection maps of spherical shells of $10h^{-1}$kpc thickness at $0.1R_{\rm vir}$, $0.5R_{\rm vir}$ and $R_{\rm vir}$. It reveals the streamers extending down from $R_{\rm vir}$ to the haloes' central regions. Also shown are the walls (seen as the threads connecting the streamers), and diffuse accretion flow. The color represents radial influx of gas mass per solid angle. Sheets existing between the streamers at larger radii dissolve on the galactic scale, i.e., $0.1 R_{\rm vir}$, which are especially detectable at $z_{\textrm f} = 6$ and $z_{\textrm f} = 4$.  

The color palette in Figure\,\ref{fig:skymap} reveals the difference in the accretion rates between the filaments and walls. The former channel up to $100\,M_\odot\,{\rm yr^{-1}\,rad^{-2}}$, while the latter less than $40\,M_\odot\,{\rm yr^{-1}\,rad^{-2}}$. Furthermore, notice how the smooth accretion (not from filaments and, or walls) comes from all directions, although at a much lower rates (few $\times M_\odot\,{\rm yr^{-1}\,rad^{-2}}$).

Figure\,\ref{fig:allfilarho} reveals that the streamers differ in their density distribution at their final redshifts $z_{\rm f}$. At lower redshifts, the gas density profiles are lower than those at higher $z$, with the decline becoming more prominent between $z_{\textrm f} = 4$ and 2, as this represents the longest time interval between different $z_{\rm f}$. At $z_{\textrm f} = 6$,  their density ranges in the interval of $\rho\sim 10^{-23}-10^{-25}\,{\textrm {g\,cm}^{-3}}$ over the distance of $400h^{-1}$kpc. At $z_{\textrm f} = 4$, this density is about a factor of 3 lower. At $z_{\textrm f} = 2$, the density is even lower and ranges between $\sim 10^{-25}-10^{-27}\,{\textrm {g\,cm}^{-3}}$.  These results are indicative of a two-fold effect: the decrease in the gas mass flux in the streamers and the overall decrease of density in the universe with a decreasing redshift. Notice, however, that no trend is observed between the high and low overdensities models in this Figure.

Figure\,\ref{fig:allfilaT} displays the radial distributions of temperature in the filamentary accreting gas (the diffuse accretion flow properties are discussed in the next section).  We distinguish a different behavior inside $R_{\rm vir}$ ($\sim 200\,  h^{-1}$kpc) and outside it, with a transition region between $0.7R_{\rm vir} - 1.3R_{\rm vir}$. Some differences can also be seen between the CW and VW models. For $z_{\rm f} = 6$ models, the temperature is rising relatively mildly inside the virial radius. At $z_{\rm f} = 4$, this rise is much steeper. Streamers at $z_{\rm f} = 2$ are hotter everywhere, and still display an additional rise towards the center. In the same way, the central temperature increases towards lower $z_{\rm f}$. 

We also observe a larger amount of hot gas inside $R_{\rm vir}$ at lower redshifts and for lower overdensity models. At higher redshifts, the gas accumulates mostly at $T\sim 10^4$\,K. The virial temperature of high-redshift haloes is higher than for the low redshift ones, which is expected as these haloes are more centrally concentrated. Moreover, the flat part of the $T$-curves broadens to $\sim {\rm few}\times 10^5$\,K at lower redshifts. 

Figure\,\ref{fig:allfilaZ} exhibits a diverse radial profile behavior of the gas metallicity. While the filamentary streamers fit in the range of $(Z/Z_\odot)\sim 10^{-4} - 1$, the radial gradients sometimes exhibit opposing trends, i.e., the metallicity can increase or decrease with radius. This can be understood when the metal production occurs in small haloes embedded within the streamers at different radii. Note that the surrounding massive haloes have been removed, but the small ones and substructures persist in these Figures. Hence, the metal contamination contributes due to the outflows and the SN feedback. We also observe that the VW models appear to be less metal rich compared to the CW models, especially in the low overdensity models. This is the result of the VW models being more gas rich due to a reduced star formation everywhere \citep{bi22a}.

Figure\,\ref{fig:allfilavr} provides the radial profiles of the inflow velocities along the filamentary streamers. The obvious trend which shows up is that these velocities increase at smaller radii but decline overall at lower redshift, i.e., become less negative inside $R_{\rm vir}$. Around 50\,kpc/h from the center they decrease sharply with decreasing $z_{\rm f}$. This effect is a direct consequence to DM halos of the same mass being less centrally concentrated, as they form in a lower density universe.  

Also, inside $\sim 100$\,kpc/h, the difference between the CW and VW becomes large, which is probably related to stronger shocks in VW models due to the larger fraction of gas in the VW models.} At $z_{\textrm f} = 6$, the inflow velocity in CW models ranges within $v_{\textrm r} \sim 100 - 300\,{\textrm {km\,s}^{-1}}$. At $z_{\textrm f} = 4$, this range becomes smaller,  $v_{\textrm r} \sim 50 - 200\,{\textrm {km\,s}^{-1}}$. While at $z_{\textrm f} = 2$ it decreases to $v_{\textrm r} \sim 70-150\,{\textrm {km\,s}^{-1}}$.  

The mass accretion rate of low-temperature gas onto the halo provides an important reservoir of gas that can join the galaxy and contributes to its SFR. Figure\,\ref{fig:allfilaflux} shows the radial profiles of the accretion rate from baryonic filaments, $\dot M$, for all models at $z_{\rm f} = 6$, 4, and 2. The individual streamer contributions have been added up. The $\dot M$ appears flat. 

The observed variations in the mass accretion rates in Figure\,\ref{fig:allfilaflux} are limited within a factor of 2 at smaller radii for $z_{\rm f} = 6$ galaxies and for high overdensity models at $z_{\rm f} = 4$, and are associated to the presence of substructures. At $z_{\rm f} = 2$, accretion rate in filaments is either flat or decreases at small radii for the high density haloes, with a slight increase for the low-density ones.

As expected, the accretion rates along the streamers decrease with decreasing $z_{\rm f}$. For $z_{\textrm f} = 6$ it lies within $\dot M\sim 50 - 100\,{\textrm M}_\odot\,{\textrm yr^{-1}}$. For $z_{\textrm f} = 4$, it ranges within $\sim 10 - 100\,{\textrm M}_\odot\,{\textrm yr^{-1}}$, and for  $z_{\textrm f} = 2$, it ranges within  $\sim 5 -10\,{\textrm M}_\odot\,{\textrm yr^{-1}}$. The spikes visible in this Figure are the result of substructure and should be ignored. 

This decrease in the accretion rates towards smaller redshifts appears in tandem with the decrease in the density within the filaments and the associated inflow velocities discussed above. We also observe this trend in the accretion rates of the diffuse gas, shown in the next section. However, we do not see any systematic differences between the accretion rates in the CW and VW models. This issue will be addressed in section\,\ref{sec:discussion4}.

In order to visualize the accretion flow on scales smaller than $50h^{-1}$kpc, Figure\,\ref{fig:surfdens} shows whole-sky HEALPix maps on scales of 5, 10, 30, 40, and $50h^{-1}$kpc, using shells of $3h^{-1}$kpc thickness and colored by the gas surface density. In this Figure, we can detect the filamentary and diffuse accretion flows, but the walls that have been observed in Figure\,\ref{fig:skymap} on larger scales, have been washed out here. The inner most panels show the gas distribution within the galaxies (aligned along the equatorial planes) and its immediate surroundings. The presence and strength of the streamers decrease at smaller radius until $R \sim 10 h^{-1}$kpc (typical radii of our galaxies) and also as a function of redshift. This situation is also independent of the wind mechanism employed. At this region, some streamers connect directly with the galaxy, but some others might miss it becoming outflows. Such interplay can be noticed by the intricate density pattern at the $10 h^{-1}$kpc shell for all models, different and almost uncorrelated with respect to their external shells.

Figure\,\ref{fig:surfdens} also shows that some of the material in the filamentary accretion misses the central galaxy, it is subsequently diverted and becomes a filamentary outflow. This is detected in the analyzed velocity field not shown here, and is typical for all the models. This outflow does not escape from the halo, reaching $\ltorder 150h^{-1}$kpc. The gas in these outflows increases its entropy and is converted into a diffuse virialized gas.  In addition, this Figure includes contribution from galactic outflows, i.e., CWs and VWs.

\subsection{Radial properties of diffuse accretion flow}
\label{sec:results422} 

Next, we turn to the radial properties of the diffuse accretion flow in the vicinity and inside the DM haloes, $\ltorder 2R_{\rm vir}$. We follow their density, temperature, radial velocity, metallicity and diffuse accretion rate. 

\begin{figure*}
    \center 
    \includegraphics[width=0.95\textwidth]{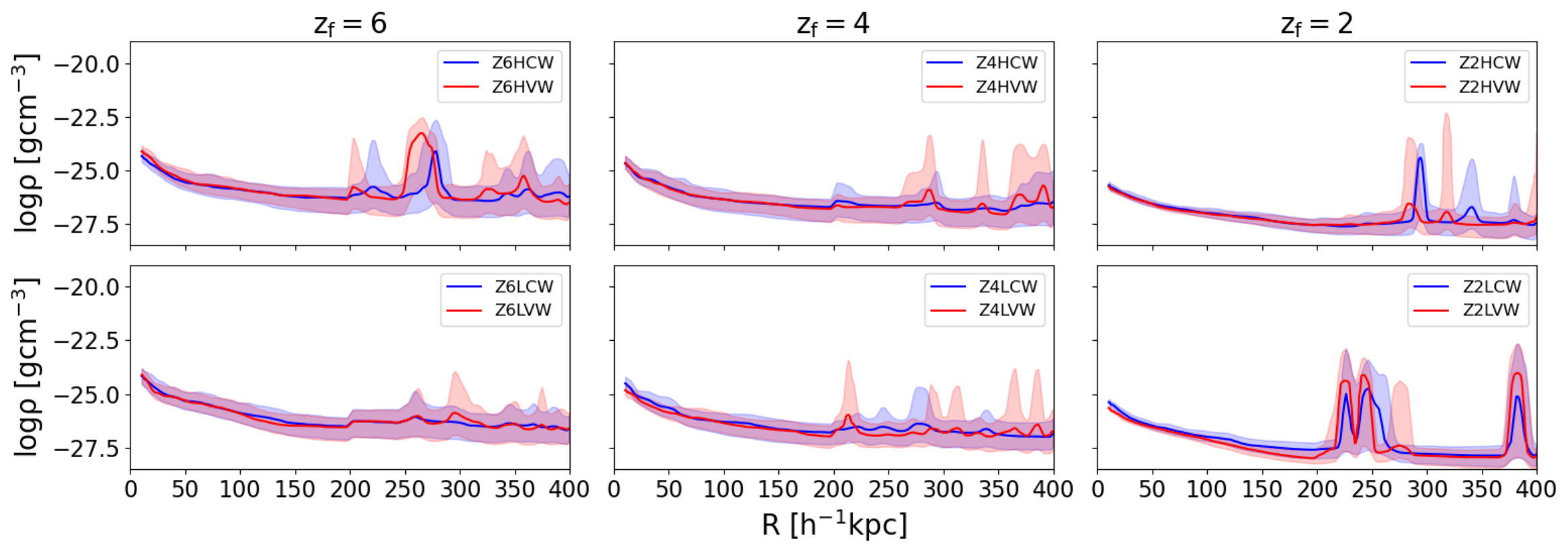}
    \caption{Radial density profiles of  diffuse gas. The VW (CW) models are given by the red (blue) lines. The color shadows represent the 20-80 percentile distributions.}
    \label{fig:alldiffuserho}
\end{figure*}    

\begin{figure*}
    \center 
    \includegraphics[width=1\textwidth]{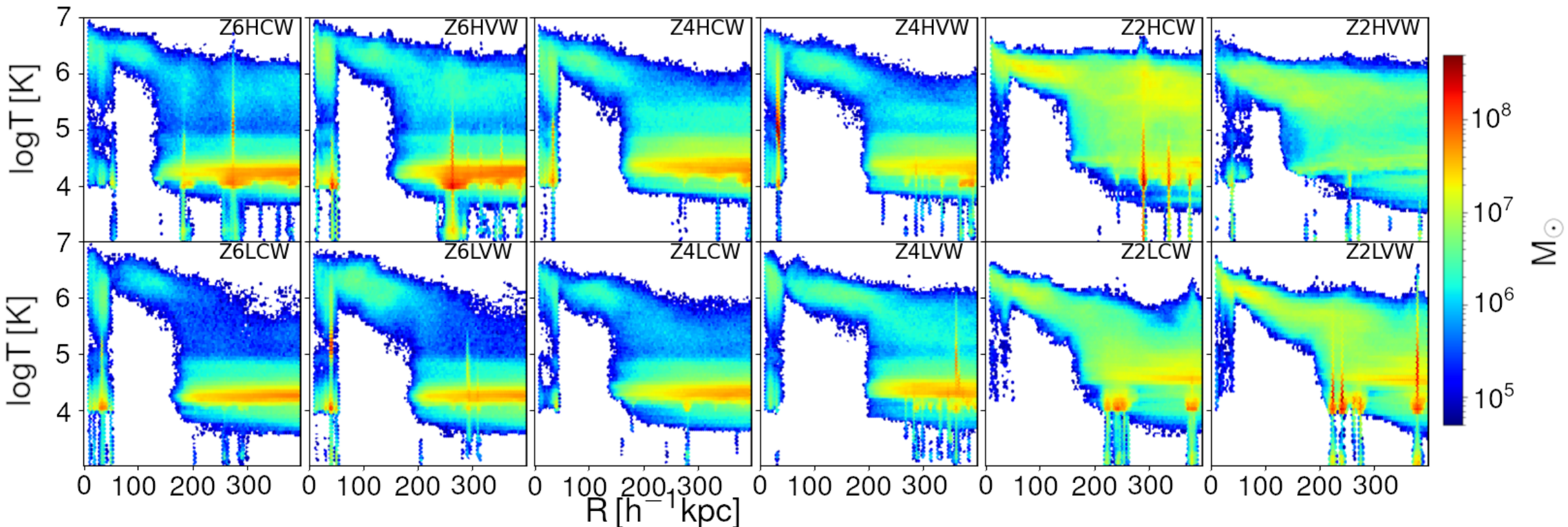}
    \caption{Hex-binned radial temperature profiles of a diffuse gas inside $2R_{\rm vir}$ with the color palette representing the gas mass in individual pixels at $z_{\rm f} = 6$, 4 and 2. Note that the virial temperatures of these DM haloes is $T\sim 1.3\times 10^6(1+z)$\,K.}
    \label{fig:alldiffuseT}
\end{figure*}  

\begin{figure*}
    \center 	\includegraphics[width=0.95\textwidth]{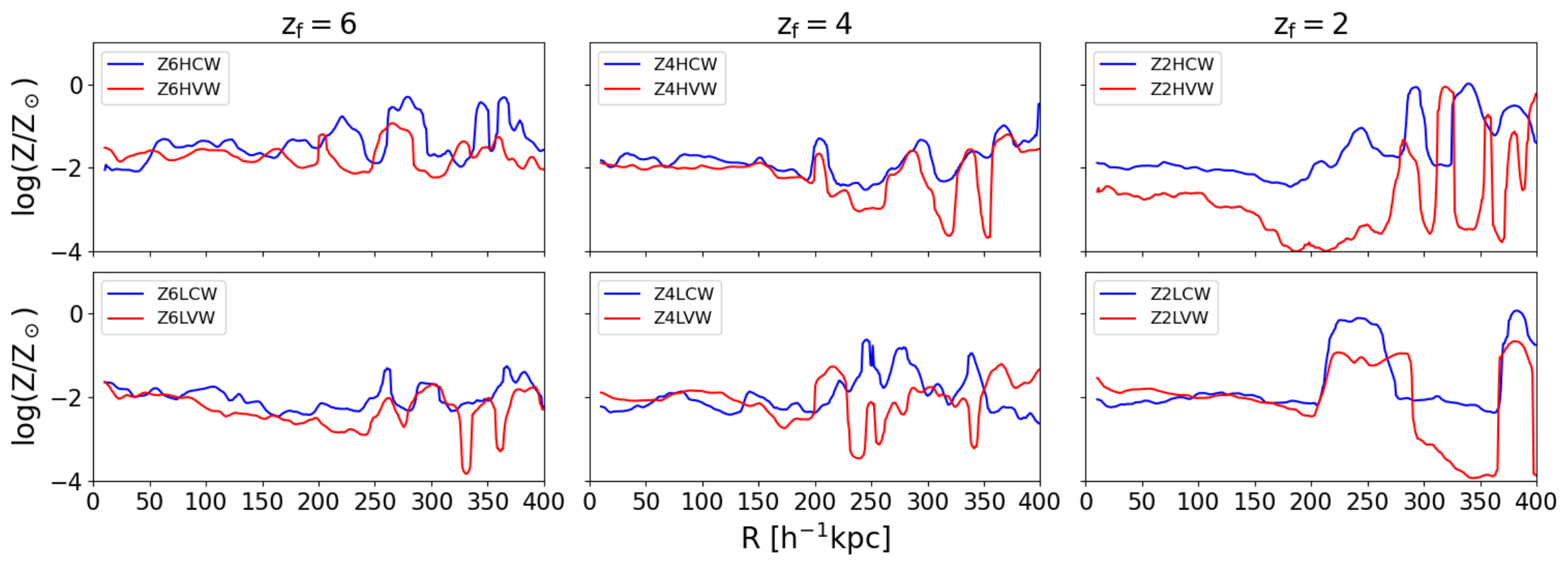}
    \caption{Mass-weighted gas metallicity radial profiles of diffuse gas. The VW models are given by the red lines, and CW models by the blue lines.}
    \label{fig:alldiffuseZ}
\end{figure*}  

\begin{figure*}
    \center 
    \includegraphics[width=0.95\textwidth]{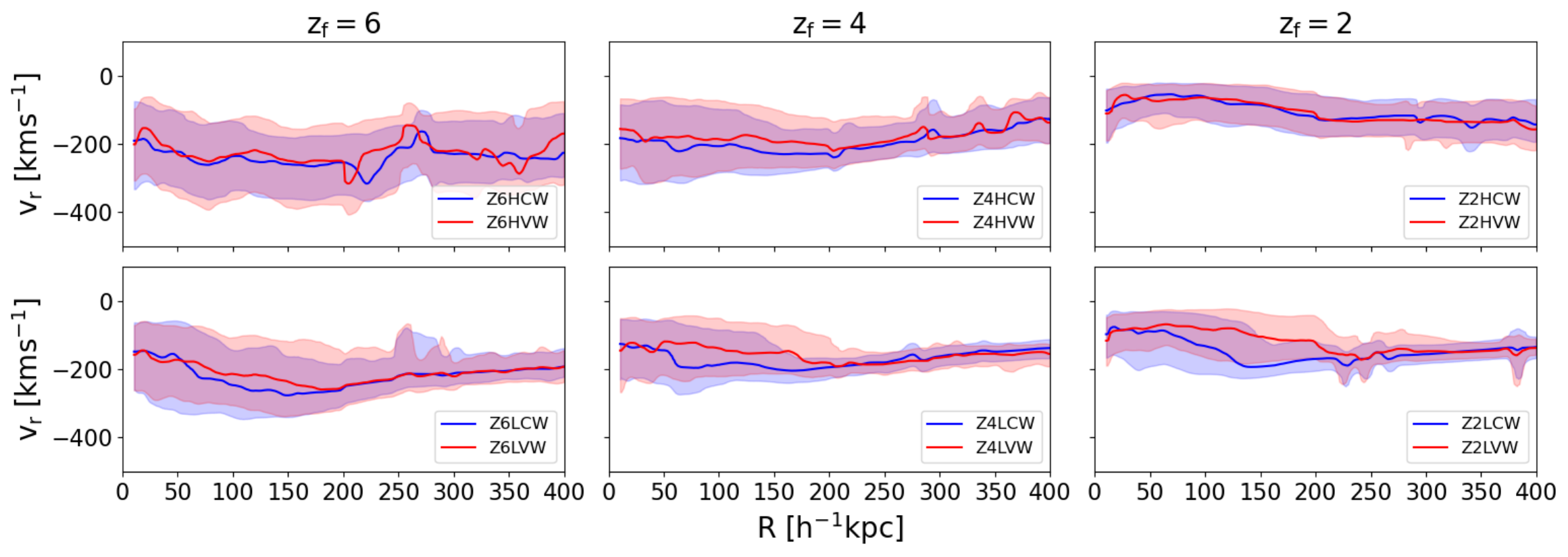}
    \caption{Radial velocity profiles of diffuse gas. The VW (CW) models are given by the red (blue) lines. The color shadows show the 20-80 percentile scale.}
    \label{fig:alldiffusevr}
\end{figure*}    

\begin{figure*}
    \center 	\includegraphics[width=0.95\textwidth]{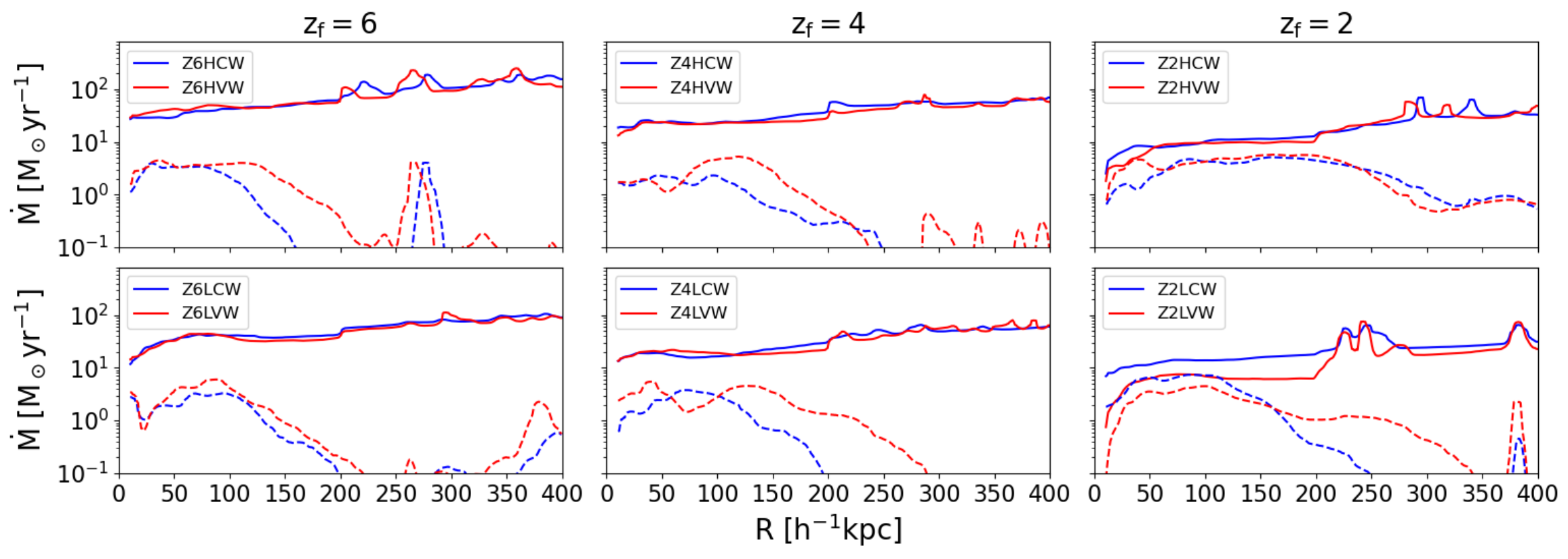}
    \caption{Mass accretion rate profiles of diffuse gas (solid lines) overplotted onto the outflows (dashed lines) triggered by feedback from the central galaxies. The VW (CW) models are given by the solid red (blue) lines. For both inflow and outflow rates, we have used the radial velocities exceeding the local dispersion velocities in the gas.}
    \label{fig:alldiffuseflux}
\end{figure*}    

Figure\,\ref{fig:alldiffuserho} shows the density profile of the diffuse accreting gas. Its radial dependence is similar to that of the filamentary gas, but shifted down by a factor of $\sim 5-20$. These profiles appear to be flat between $(1-2)R_{\rm vir}$ and $\rho\sim R^{-1}$ inside the virial radius.  The explanation to this behavior is similar to that in the filamentary accretion mentioned in section\,\ref{sec:results421} --- decrease in the gas supply as well as the expansion of the universe, both contribute to this effect. We observe no differences between the density profiles based on environment neither with respect to wind model. 

The temperature in the diffuse accretion gas (Figure\,\ref{fig:alldiffuseT})  differs from that in the filaments (Figure\,\ref{fig:allfilaT}). The lion share of the gas mass in the filaments has a temperature of $\sim 10^4$\,K, except for $z_{\rm f} = 2$. The diffuse gas however, has such a low temperature only outside the virial radius. At smaller radii, the majority of the gas lies at $T\sim {\rm few}\times 10^5 - 10^7$\,K. The cold diffuse flow is ceased to exist inside $R_{\rm vir}$. At $z_{\rm f} =2$, the amount of hot gas outside $R_{\rm vir}$ is also increased. Some differences among the models at $z_f$ regarding their local overdensities can be noticed, in particular the amount of diffuse gas around and beyond $R_{\rm vir}$, it is much larger the amount of diffuse gas for the high-density environment haloes (as expected) with respect to their low-density counterparts.  In summary, the thermodynamic state of the gas in the filamentary and diffuse accretion differ, and this can have implications for the gas kinematics as well, due to the buildup of thermal pressure gradients.

Figure\,\ref{fig:alldiffuseZ} exhibits the mass-weighted metallicity radial profiles of the diffuse accreting gas. The profiles are relatively flat with significant variations related to the embedded substructures and small neighboring haloes. Simulations with the VW feedback exhibit a lower metallicity than those with CW feedback, especially outside the virial radius. This is mainly due to the overall lower SFR in these models when compared to the CW ones. Furthermore, this is quite similar when compared with the filamentary accretion flows (Figure\,\ref{fig:allfilaZ}). Both distributions display a substantial fraction of low metallicity gas, $Z/Z_\odot \ltorder 10^{-3}$ and even pristine gas outside $R_{\rm vir}$.  

The radial inflow velocity profiles of the diffuse accreting gas, while following the same trend as the filamentary accretion in general, differ from the gas flow in the filaments in detail (Figure\,\ref{fig:alldiffusevr}). Namely, the velocities of the diffuse gas appear smaller at lower $z_{\rm f}$, but their radial velocity gradient is not as large as in the filaments shown in Figure\,\ref{fig:allfilavr}. Overall, as in the filamentary accretion, the diffuse gas accretes in more concentrated halos at smaller redhift. But the differences in the thermodynamic state between the diffuse and filamentary gas lead to variations in the innermost inflow velocity profiles.

Next, we calculate the mass accretion rates, $\dot M$, of the diffuse gas (Figure\,\ref{fig:alldiffuseflux}), and separate it from the outflows triggered by the feedback from the central galaxies. To distinguish the inflow and outflow from the virialized gas inside the DM haloes, we have counted only the inflow and outflow velocities which exceed the local gas velocity dispersions. The resulting mass accretion rates for the diffuse gas decrease with approaching the central galaxies. They also decline with decrease of $z_{\rm f}$ from $\dot M\sim 10-40\,M_\odot\,{\rm yr^{-1}}$ at $z_{\rm f} =6$ to $\sim 5-10\,M_\odot\,{\rm yr^{-1}}$ at $z_{\rm f} = 2$. We do not see differences between the CW and VW models, except at the $z_{\rm f} = 2$ low overdensity Z2L model.   

The outflows are clearly focused on the central galaxies and become more extended, i.e., reaching larger radii, at lower $z_{\rm f}$. At $z_{\rm f} = 6$, the outflows reach $\sim R_{\rm vir}$, at $z_{\rm f} = 4$, the VW outflow is reaching $\sim 1.5R_{\rm vir}$, while at $z_{\rm f} = 2$, in the low-density environment haloes the CW reaches $\sim 1.3R_{\rm vir}$ while VW extends to $\sim 1.8R_{\rm vir}$. In general, the outflows seen at larger radii belong to small haloes located well beyond $R_{\rm vir}$. 

The outflow radial profiles decrease with radius. At the maximum, the outflows are $\sim 5-10\,M_\odot\,{\rm yr^{-1}}$ at all $z_{\rm f} =6$, but decline below $0.1\,M_\odot\,{\rm yr^{-1}}$ at the maximal radii mentioned above. But in all cases, we find that the outflow extends to the virial radius, at least. In addition, some of the inflow misses the central galaxy and is diverted out entirely because of gravitational effects. It interacts with the inflow even outside $R_{\rm vir}$ as we discuss in section \ref{sec:discussion4}. Based on this result, we call this region, which extends from the HOP-defined galaxy to the baryonic backsplash radius as the CGM. Thus, the CGM is affected both by outflows and inflows, and, therefore, has a complicated kinematics and other thermodynamic characteristics. 

\subsection{Dissecting the streamers: emerging spaghetti-type flow} 
\label{sec:results43}

While analyzing the radial profiles of various thermodynamic properties of streamers at specific times is necessary for obtaining the global picture, following the time-dependent processes is required as well \citep[e.g.,][]{arz11}. As a next step, we investigate the radial motions within the streamers in the range of $\sim 50 - 350\,h^{-1}$\,kpc from the central galaxy, and determine their structure at different radii by dissecting them into slices of $10\,h^{-1}$\,kpc radial thickness. This is complemented by the next section (section\,\ref{sec:results44}), where we focus on the inner $50\,h^{-1}$\,kpc, where the inflow gradually dissolves, changing its kinematics before penetrating the central galaxy. 

Analysis of the streamers radial profiles and their cross section structures at different radii is important for our understanding of their evolution within the DM haloes. It also reveals the fate of these streamers at radii comparable to galaxy sizes. As a prototype, we use the streamer \#1 of the Z2L halo identified in Figure\,\ref{fig:allfil}, and analyze it down to $\sim 10\,h^{-1}$\,kpc. For this purpose, we select the gas particles outside $R_{\textrm {vir}}$ at $z\sim 2.4$ and trace them inwards until $z_{\rm f} = 2$. This streamer falls into the central galaxy being inclined with respect to the disk plane by an angle of $\sim 30^\circ$.

\begin{figure}
\center 
    \includegraphics[width=0.5\textwidth]{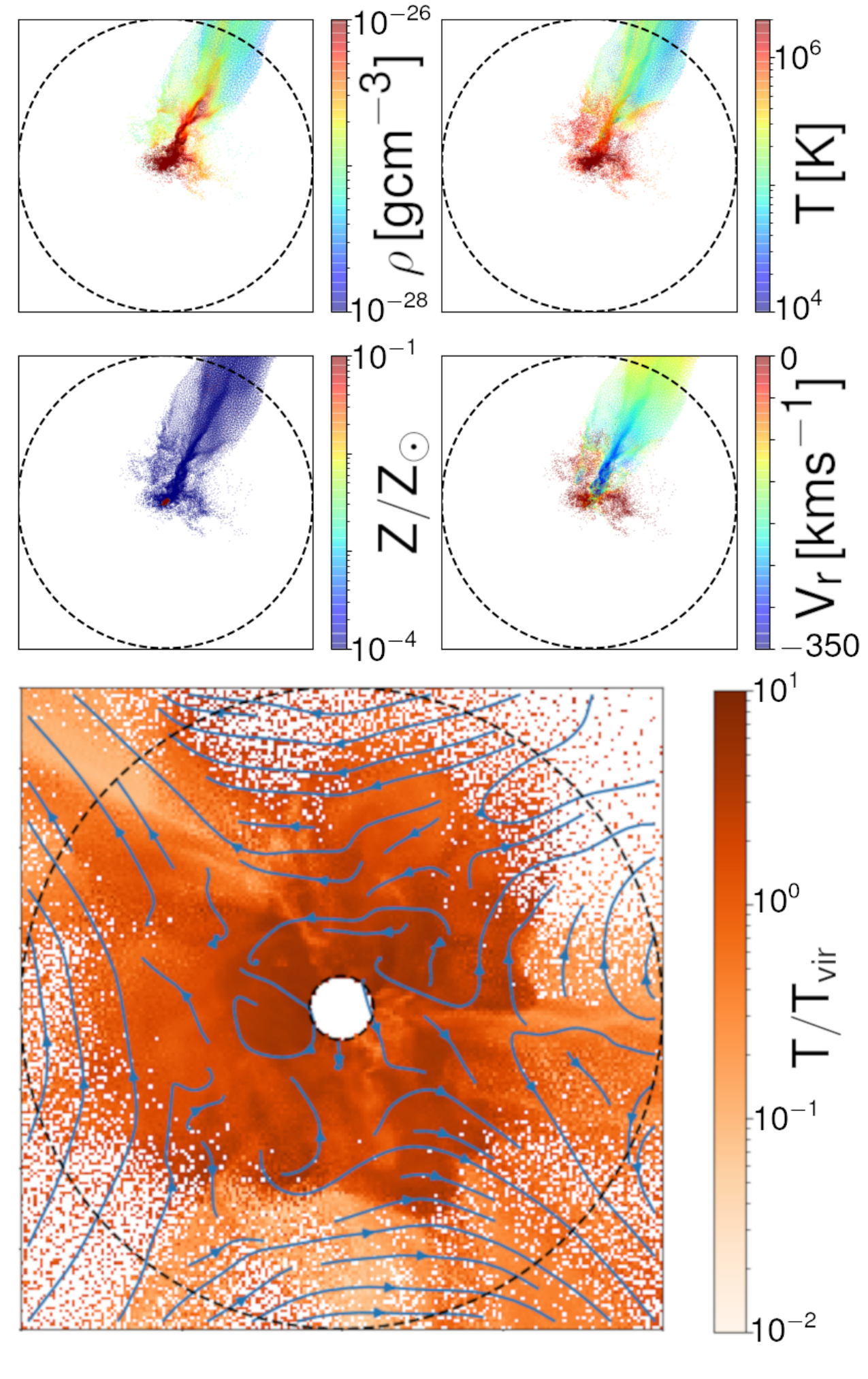}
    \caption{{\it Upper plot:} the gas particles in the filamentary streamer \#1 of the halo Z2L at $z = 2.1$ (see Figure\,\ref{fig:allfil}), colored with density (top left), temperature (top right), metallicity (bottom left) and the infall velocity (bottom right). {\it Lower plot:} the CGM temperature in a slice with $0.1 R_{\rm vir}$ thickness, normalized by $T_{\rm vir}$ from 0.1 (galaxy's radius, inner circle) to $R_{\rm vir}$ for the halo Z2L. The velocity field is traced by the blue streamlines.  
    }
    \label{fig:fila1edgeon}
\end{figure}

Properties such as the volume density, temperature, metallicity and the radial velocity of this gas show an interesting structure as seeing in Figure\,\ref{fig:fila1edgeon} (upper plot). Most prominent is the inner, central region of the streamer, i.e., the radial spine, at almost all radii, where one can define the {\it core} of the streamer for each of these variables. We also observe that the above physical properties correlate along the streamer. For example, the volume density and the temperature show a tight correlation. However, we do not see a gradient of metallicity between the core and the envelope of the streamer.  

\begin{figure*}
\center
    \includegraphics[width=0.98\textwidth]{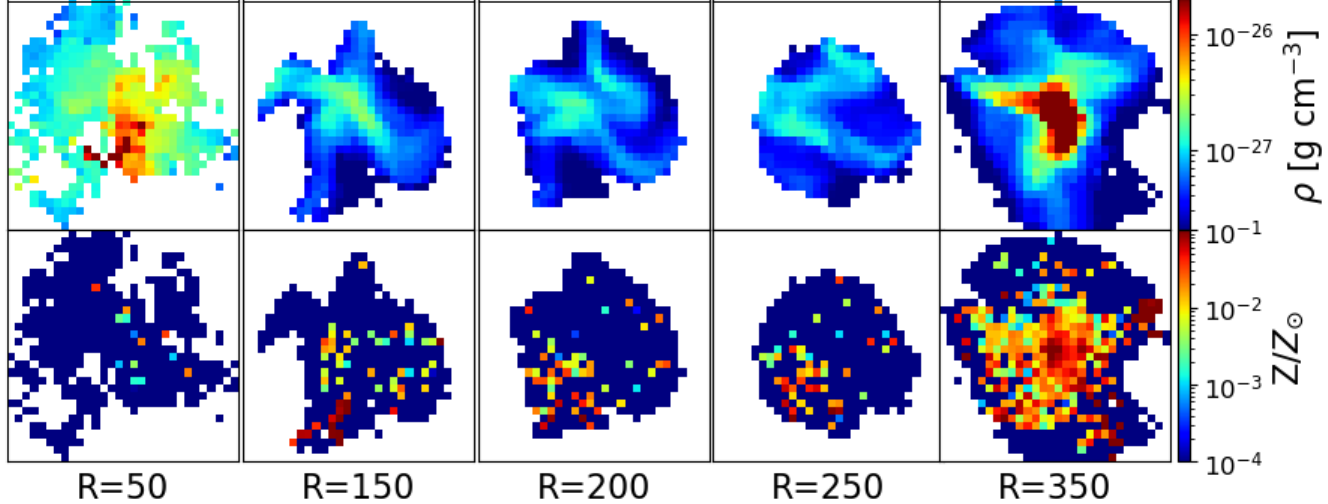}
    \caption{Representative slices of the gas in the filament \#1 of halo Z2L (see Figure\,\ref{fig:allfil}) at $z_{\textrm f} = 2$. Shown the gas density (top) and metallicity (bottom). From left to right the slices represent the cuts at distances $R =  50$ (the inner CGM), 150 (the outer CGM), 200, 250, and $350\,h^{-1}$\,kpc. Each slice has a $10\,h^{-1}$\,kpc radial thickness. Each frame has $120\,h^{-1}$\,kpc length on the side.  
    }
    \label{fig:filacross}
\end{figure*}

To verify this, we have dissected this streamer perpendicularly to its spine. Figure\,\ref{fig:filacross} provides a more revealing view to study the gradients between the core and the envelope of the streamer, and their radial dependence. For the volume density, the cross section at the innermost radius $R\sim 50\,h^{-1}$\,kpc (in the midst of the inner CGM) displays a high density core of $\rho\sim 10^{-25.5}\,{\rm g\,cm^{-3}}$, while somewhat less denser at 150 (in the midst of the outer CGM), 200 (the virial radius), 250, and $350\,h^{-1}$\,kpc. The envelope density of the streamer drops by more than one order of magnitude compared to that at the inner boundary of $R\sim 50\,h^{-1}$\,kpc. The core density has a weaker dependence on radius than the envelope.  

Small scale structures and substructures are typically embedded in the streamer. For example, the slice situated at $R\sim 350\,h^{-1}$\,kpc of Figure\,\ref{fig:filacross} (i.e., the lowest frame)  displays a structure embedded in the spine of the streamer. This trapping of the structure in the core of the streamer explains partly its very high density. The temperature minimum typically lies within the streamer's core, this is also the case for this snapshot due to the cold gas contained within the structure.   

Most of the slices exhibit almost pristine metallicity, $(Z/Z_\odot)\sim 10^{-4}$. But high-metallicity gas particles of $(Z/Z_\odot)\sim 10^{-3} - 10^{-1}$ are found scattered across the cross-section without a strong correlation with the density or temperature. This is in agreement with the metallicity radial profile in Figure\,\ref{fig:fila1edgeon}. At the same time, the region surrounding the structure always has a high metallicity, and this is confirmed by the slice at $R\sim 350\,h^{-1}$\,kpc.

We have measured the radial infall velocity at the innermost radius of $R = 50\,h^{-1}$\,kpc, and it has reached $v_{\textrm r} \sim 250 - 200\,{\textrm {km\,s}^{-1}}$ for the core gas. But the infall velocity for the envelope drops to as low as $v_{\textrm r} \sim 25-50\,{\textrm {km\,s}^{-1}}$. At larger radii, the infall velocity gradually decreases from the core to the envelope from $v_{\text r} \sim 200\,{\textrm {km\,s}^{-1}}$ to below $v_{\text r} \sim 50\,{\textrm {km\,s}^{-1}}$. Such velocity gradients will induce shear between the core and the envelope, and trigger turbulence. Furthermore, the high density zones, i.e., the cores,  tend to separate into different infall velocity regions with a spaghetti-like morphology. Both the developed turbulence and the spaghetti-type flow are characteristics of dissolution of the filamentary flow and its gradual virialization.

The lower plot of Figure\,\ref{fig:fila1edgeon} displays the 0.1$R_{\rm vir}$ thick slice of the same halo, Z2L, with the superposed velocity field. The kinematics of filamentary streamers is delineated here, supplemented by increasingly virialized gas around the central region. The major streamers can be seen partly hitting the central galaxy region and partly missing it converting the inflow into outflow. The emergence of the central turbulent region can be observed as well. Clearly, the gas withing the halo is not in hydrostatic or thermal equilibrium.

\subsection{Joining the central galaxy: dissolution of the accretion streamers} 
\label{sec:results44}

\begin{figure}
\center 
    \includegraphics[width=0.50\textwidth]{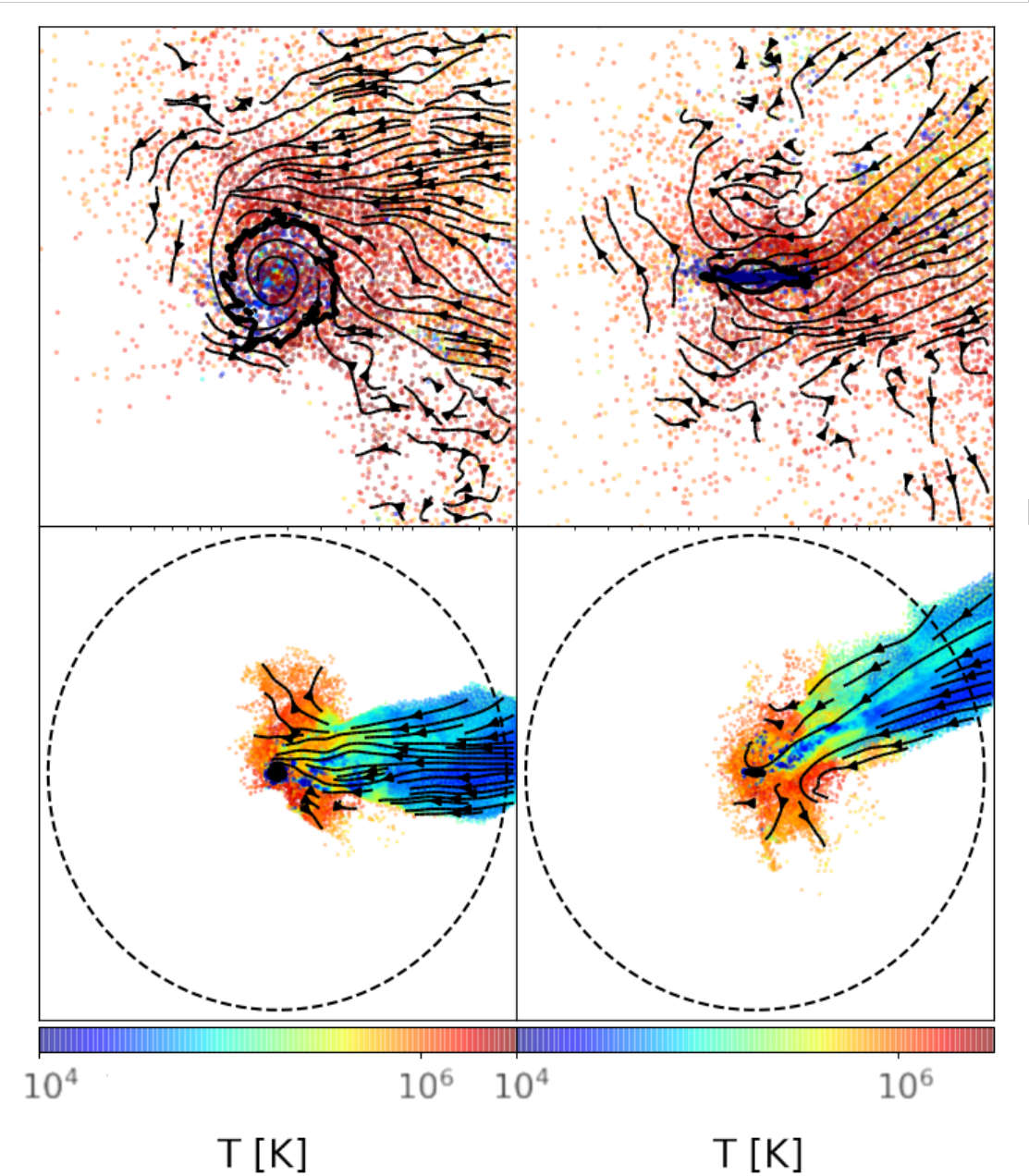}
    \caption{Face-on (left) and edge-on (right) maps of the gas stream in filament \#1 of the halo Z2L at $z_{\textrm f} = 2$ (see Figure\,\ref{fig:allfil}), colored with the gas temperature. The upper panels represent a region of the $60\,h^{-1}$\,kpc side,  and the lower panels of $400\,h^{-1}$\,kpc side. The dashed circles represent the virial radii. The central black contours in the upper boxes display the shape of the central galaxy, i.e., face-on and edge-on. }
    \label{fig:fila1join2}
\end{figure}  

\begin{figure}
    \center
    \includegraphics[width=0.48\textwidth]{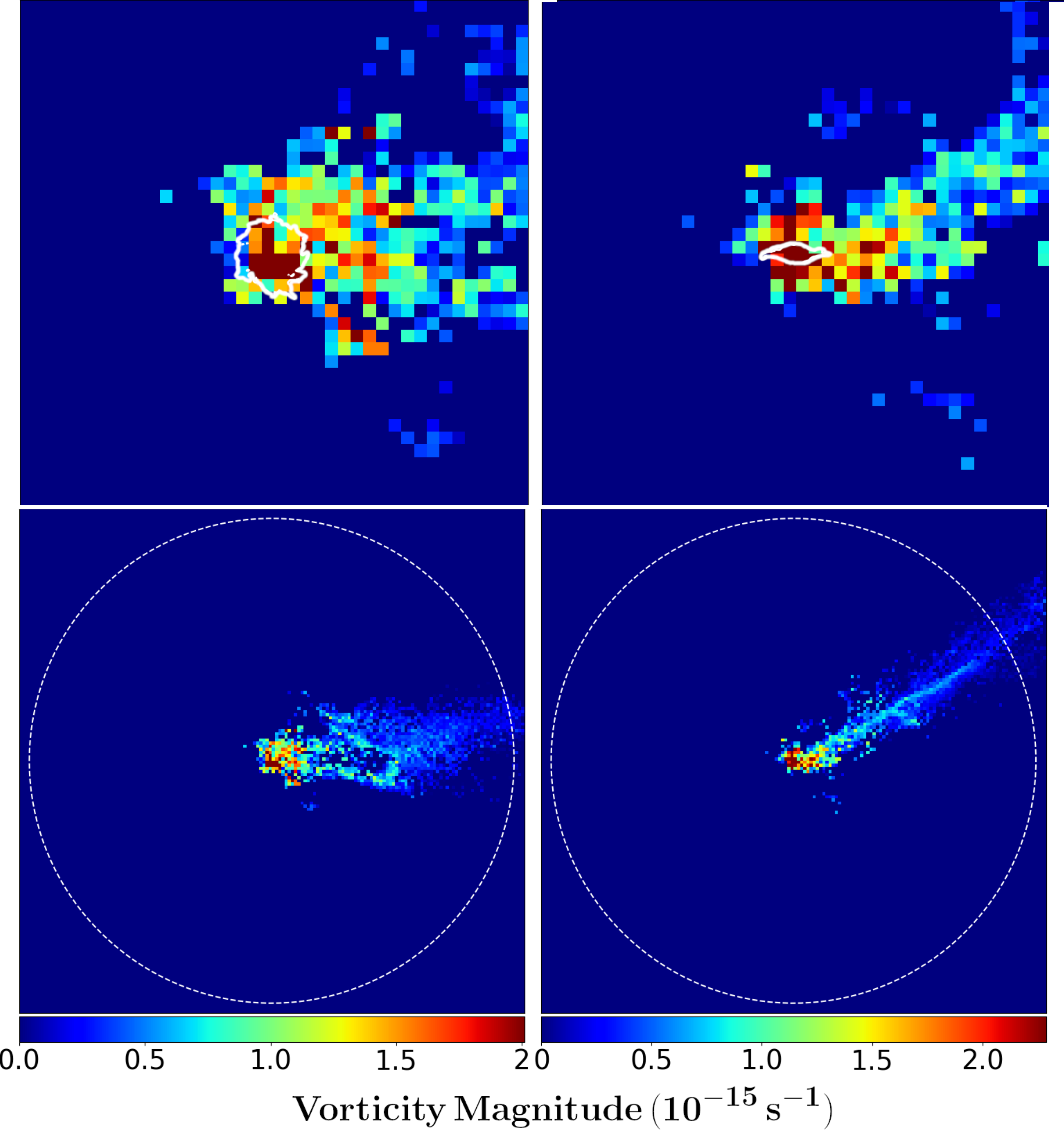}
    \caption{Face-on (left) and edge-on (right) vorticity maps of the gas particles in the filament \#1 of the halo Z2L at $z_{\textrm f} = 2$ (see Figure\,\ref{fig:allfil}). The upper panels shown in the $60\,h^{-1}$\,kpc boxes and the lower panel shown in the $400\,h^{-1}$\,kpc boxes. The dashed circles represent the virial radii. The central white contours in the upper frames display the shape of the central galaxy, i.e., face-on (left) and edge-on (right). }
    \label{fig:fila1join}
\end{figure}

\begin{figure*}
    \center	\includegraphics[width=0.95\textwidth]{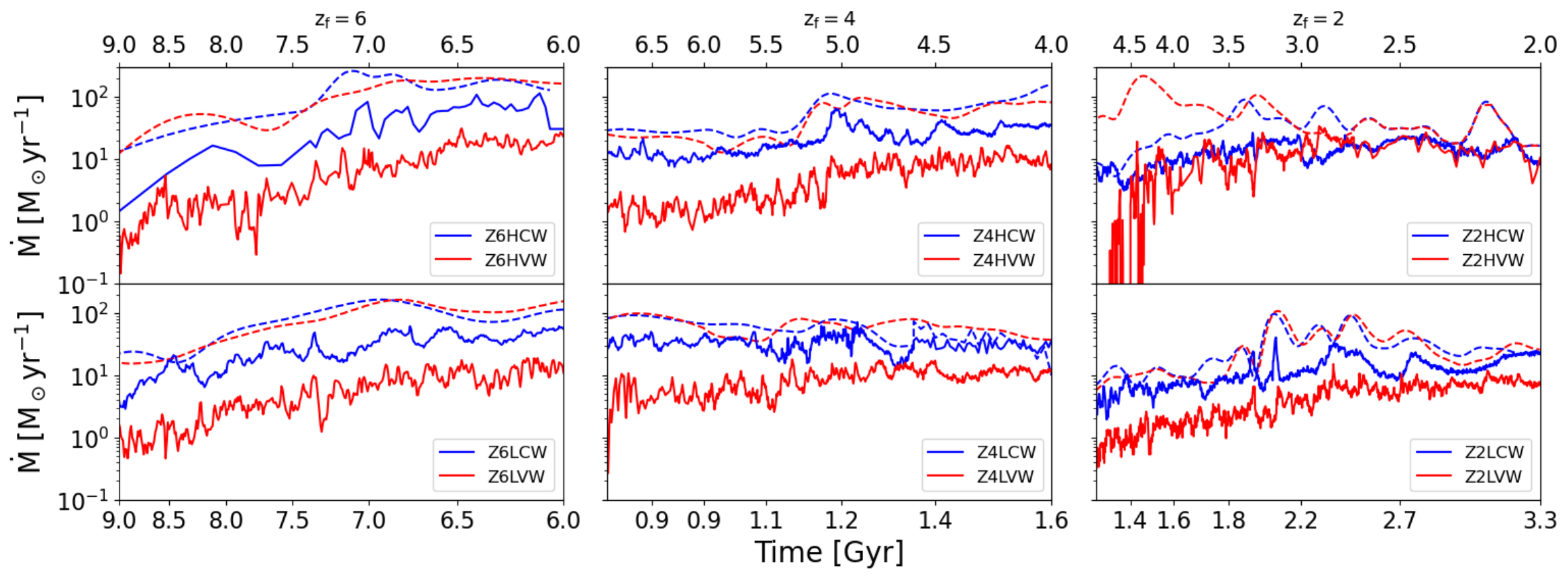}
    \caption{Evolution of the SFR (solid lines) and galaxy growth rates (dashed lines) in all models of our simulation suite. The CW and VW models are represented by blue and red lines, respectively. The galaxy growth rate (gas $+$ stars) corresponds to the baryonic accretion rate minus the outflow rates.  The galaxies residing in a high density environment are shown in the upper panels and those in low density environment lie in the lower panels. }
    \label{fig:fila2join}
\end{figure*}

\begin{figure*}
    \center
    \includegraphics[width=0.95\linewidth]{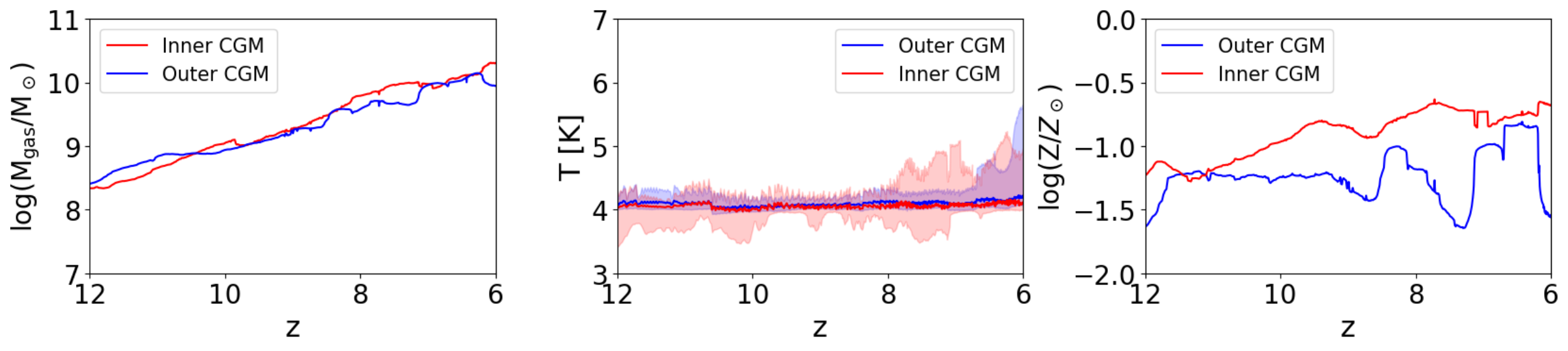}
    \caption{Evolution of the CGM of the Z6LCW halo model. {\it Left:} the gas mass in the inner (red line) and outer (blue line) CGM, which envelops the central galaxy and extends to $\sim R_{\rm vir}$. The inner CGM is defined within $0.5R_{\rm vir}$, and the outer CGM lies in the range of $0.5R_{\rm vir}-R_{\rm vir}$; {\it Center:} the median temperature evolution of the inner (red) and outer (blue) CGM. The shaddows represent the 20-80 percentiles. {\it Right:} average metallicity evolution of the inner (red) and outer (blue) CGM.}
    \label{fig:awesome_image3}
\end{figure*}

\begin{figure}
    \center 
    \includegraphics[width=0.5\textwidth]{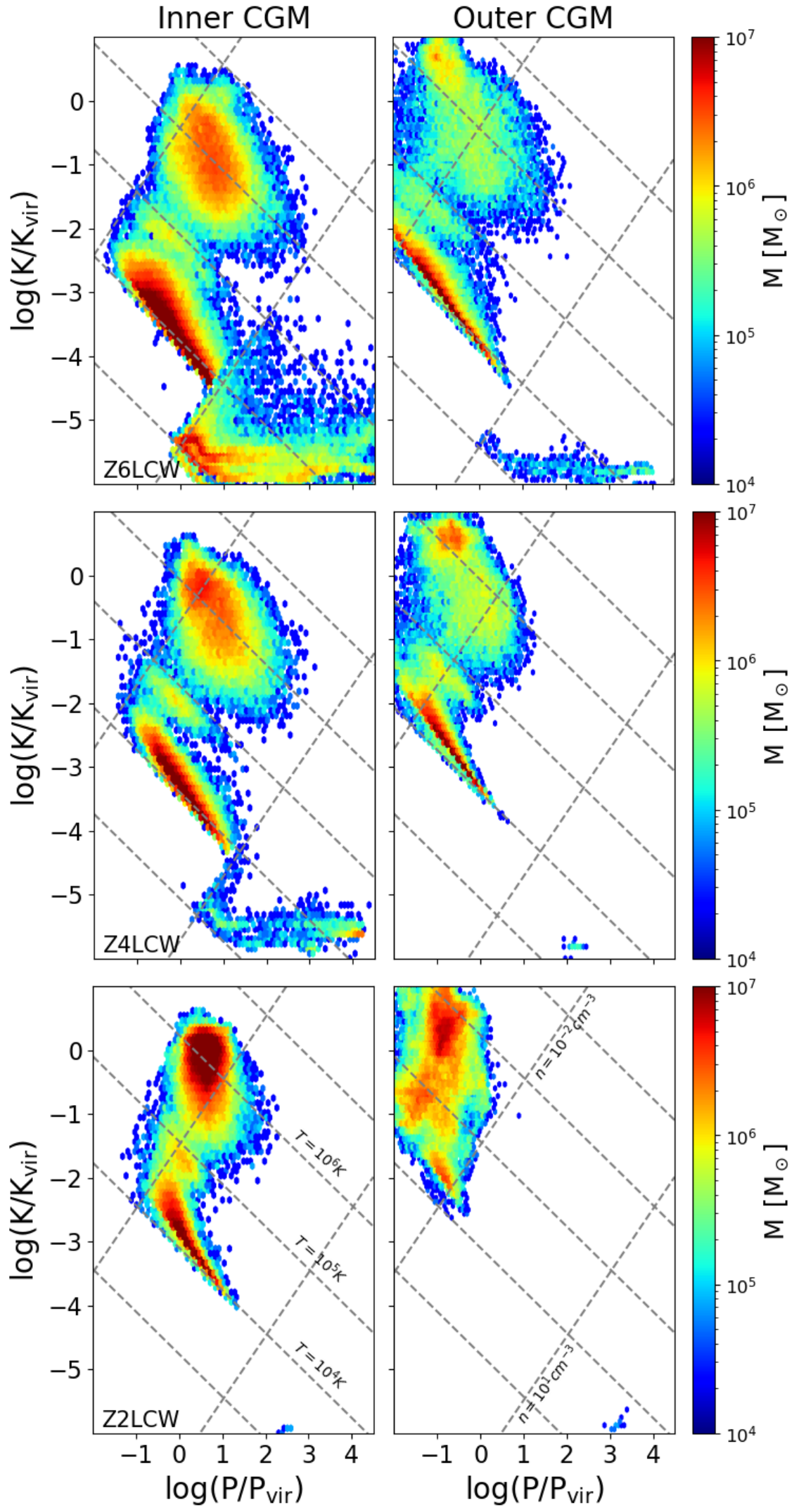}
    \caption{Pressure–entropy phase diagram in $0.1 R_{\rm vir}$ thick shells in the inner (centered on $0.25 R_{\rm vir}$) and outer (centered on $0.75 R_{\rm vir}$) CGM of three haloes LCW at $z_{\rm f} = 6$, 4 and 2. The dashed lines show lines of a constant temperature ($10^3-10^7$\,K) and constant density ($10^{-3}-10^1\,{\rm cm^{-3}}$). Note that the color palette is given in $M_\odot$.  }
    \label{fig:pressure}
\end{figure}

We proceed to address the evolution of the streamers in the innermost halo region of $R\sim 10 - 50\,h^{-1}$\,kpc, where it is more difficult to separate the streamers from the smooth accretion and where the kinematics becomes more complex --- thus characterizing the virialization process of the streamers. This region is the most interesting one because it borders with the HOP-defined central galaxy --- we call it the {\it inner} CGM in contrast to the overall CGM which extends to the baryonic backsplash radius. 

As stated earlier, in order to analyze the streamer's kinematic and thermodynamic properties, we identify their particles at larger radii and follow them individually to the center. For the sake of not overestimating the inflow rate within this region, we only consider particles belonging to the streamer or to the diffuse flow if their radial velocity exceeds the local dispersion velocity in the gas.

Figure\,\ref{fig:fila1join2} displays the fate of the streamer \#1 within the halo Z2L discussed earlier (see also Figure\,\ref{fig:allfil}), using two scales: a small scale of $30\,h^{-1}$\,kpc (top panels) and a large scale representing the virial radius, $\sim 200\,h^{-1}$\,kpc (bottom panels). On both scales the streamer and the flow streamlines are presented in two perpendicular projections with respect to the central galactic disk, face-on (left column) and edge-on (right column). We noted before that this streamer approaches the galaxy with an angle of $\sim 30^\circ$ to the galactic equatorial plane. In fact, in our models, the streamers appear to be inclined to the galaxy planes at various angles, as has been shown in previous simulations \citep[e.g.,][]{hell07,shlo13,krol19}. 

In the lower panel of Figure\,\ref{fig:fila1join2}, the gaseous streamer can be seen maintaining the temperature around $T\sim 10^{4}-10^{5}\,{\textrm K}$ when falling through the virial radius, and it is heated up dramatically to $T\sim 10^{6}\,{\textrm K}$ at around $R\sim 50\,h^{-1}$\,kpc. The streamer has a thickness, i.e., diameter, of $\sim 120\,h^{-1}$\,kpc. This is much wider than the galaxy scale of $\sim 20\,h^{-1}$\,kpc. Most of the gas falls towards the center being aligned with the direction of the streamer until $R\sim 30\,h^{-1}$\,kpc, where the gas streamlines become distorted and more turbulent (see also upper panels). Some of the infalling gas which missed the galaxy is converted into an outflow here, which extends to the {\it splash} radius --- the first apocentric passage, which in most of the cases lies outside $R_{\rm vir}$. Although originally defined with respect to  DM \citep[e.g.,][]{diemer14,adhikari14}, it has also a baryonic feature formed by the caustic generated by the piling up of the accreted material near apocenters.

Analysis of a turbulent motion is inherently difficult. We follow the method outlined in \citet{choi13}, which relies on the vorticity and its cross product with the velocity field, $\vec{w}=\bigtriangledown \times \vec{v}$. We have pixelized the filament image and plotted the vorticity field in Figure\,\ref{fig:fila1join} at the galaxy (upper frames) and the halo scales (lower frames) as specified above. On large scales and outside the streamer, the flow appears to be irrotational. Inside the streamer, the vorticity increases closer to the galaxy. This is partly due to the shear discussed above and the resulting Kelvin-Helmholtz instability which triggers ablation of the streamer gas inside $R\sim 30\,h^{-1}$\,kpc. Ablated gas contributes to the turbulent media surrounding the galaxy. Understanding and quantifying this ablation process is of a prime importance for the evolution of streamers at smaller radii.

Figure\,\ref{fig:fila2join} displays the growth rate of galaxies (dashed lines) in our sample overplotted on their SFRs (solid lines). For $z_{\rm f}=6$ models, the growth rate exceeds the corresponding SFR by about an order of magnitude. For $z_{\rm f}=4$ the difference between the growth rate and the SFR here decreases towards $z_{\rm f}$, independently of the environment and feedback. The decrease comes clearly because of the reduced growth rate, which depends on the net accretion rate.  

\section{Discussion}
\label{sec:discussion4}

We used high-resolution zoom-in cosmological simulations to trace the structure and evolution of the cosmological gas streamers penetrating DM haloes until the galaxy boundaries at selected final redshifts of $z_{\textrm f} = 6$, 4, and 2. All the haloes have been chosen to have similar masses of ${\rm log}\,M_{\textrm {vir}}/{\textrm M_\odot}\sim 11.75\pm 0.05$ at their final redshifts, and evolving in high or low density environments, i.e., in different overdensities. Furthermore, the resulting central galaxies have been subjected to different types of feedback. We compared the thermodynamic and kinematic properties of streamers and diffuse accretion at these redshifts, starting outside the virial radii and down to the central galaxy regions. We analyze the dissolution process of streamers due to their interaction with the galactic environment extending to the halo virial radius and forming the CGM --- the gaseous counterpart of DM halo.

We start by summarizing our results and analyze them. Specifically, we find that, 

\begin{itemize}

\item Using a hybrid d-web/entropy method applied to the gaseous filaments, i.e., streamers, allows us to map these streamers down to $\sim 10\,h^{-1}$\,kpc, i.e., down to the central galaxy scales. Applying this method in tandem with the gas kinematics provides an efficient way to separate the inflow from outflow and from the virialized gas within the parent DM haloes.

\item Accretion rates decrease with decreasing final redshifts, whether they proceed via streamers or via diffuse accretion --- this is a direct consequence of the reduction of available gas and expansion of the universe. The typical density of both types of accretion declines with redshift as well, i.e., from  $\rho\sim 10^{-24}-10^{-25}\,{\textrm {g\,cm}^{-3}}$ at $z_{\textrm f} = 6$ to $\sim 3\times 10^{-26}-10^{-27}\,{\textrm {g\,cm}^{-3}}$ at $z_{\textrm f} = 2$. We do not observe a systematic difference in the accretion rates in CW and VW models, not in the filamentary or diffuse modes. This result needs to be explained, and we address it in this section.

The temperature inside the streamers increases inside the virial radii, and faster at lower redshifts. However, when mass-weighted, it shows that majority of gas remains at low temperature, $\sim 10^4$\,K, at $z_{\rm f}=6$ and 4, and only at $z_{\rm f}=2$ it spreads more equally in mass for $T\sim 10^4-10^6$\,K. The temperature of the diffuse mode of accretion is higher than in the streamers inside the haloes, and while most of the gas is at $\sim 10^4$\,K outside the virial radius, it forms a bi-modal distribution of $T\sim 10^{4.5-5}$\,K and $\sim 10^{5.5-7}$\,K.  In the presence of a wide range of temperatures and velocities there, a wide range of ionization is expected as well. The maximal radial velocities within the streamers decrease with a decreasing redshift, from $\sim 200-300\,{\textrm {km\,s}^{-1}}$ at $z_{\rm f}=6$, down to $\sim 100-150\,{\textrm {km\,s}^{-1}}$ at $z_{\rm f} = 2$. The infall velocities for the diffuse gas are smaller compared to the streamers, due to the difference in the thermodynamic state of the diffuse and filamentary gas, as we have discussed in section\,\ref{sec:results422}. 

\item Outside $\sim 2R_{\rm vir}$, both the streamers and the diffuse gas accretion have metallicity in the range of $Z/Z_\odot\sim 10^{-4} - 10^0$, while inside the haloes, the metallicity distributions are flat.  The VW models appear less metal rich than CW models, due to the larger gas fraction in VW galaxies and lower SFRs. The radial inflow velocities inside the central $\sim 100h^{-1}$kpc  are typically lower for VW models than CW, possibly due to the stronger shocks in the former.  

\item Within the DM haloes, the streamers exhibit a core-envelope structure: the core radial flow surrounded by a lower density envelope. The core possesses an elevated density and lower temperature, a similar metallicity to that in the envelope, but a higher inflow velocity. This velocity gradient between the core and its envelope induces shear. At smaller radii, inside $\sim 50h^{-1}$\,kpc, the core-envelope structure is fading away and the filamentary flow splits and can be described as a spaghetti-type flow. Within the central $\sim 30\,h^{-1}$\,kpc of the streamers, the shear triggers the Kelvin-Helmholtz instability and turbulence which contribute to the ablation process of the streamers and dissolves them. We have quantified this turbulence and mapped it in two projections.

\item Galactic outflows in all our models reach the characteristic distance of $\sim 0.5R_{\rm vir}\sim 100h^{-1}$kpc from the central galaxies. The maximal  outflow rates are located close to those galaxies, $\sim {\rm few}\times M_\odot\,{\rm yr^{-1}}$, which is less than 10\% of the inflow rate at $z_{\rm f} = 6$ and 4. But for $z_{\rm f} = 2$ models, the accretion rates have declined and the maxima of the galactic outflows have increased. Hence they differ by factor of $\sim 2$ only. Moreover, the outflow rates decline substantially with the distance to the galaxy, to $\ltorder 0.1\,M_\odot\,{\rm yr^{-1}}$ at $\sim R_{\rm vir}$. 

\item The entire DM halo region has a complex kinematics involving filamentary and diffuse accretion, and galactic outflows. Some of the filamentary accretion is deflected and extends back to the baryonic backsplash radius, $\sim  R_{\rm vir}-2R_{\rm vir}$. Based on the modeled properties of such a multiphase CGM, one must conclude that it cannot be in thermal equilibrium, and its dynamic and thermodynamic state are strongly time-dependent.

\item We do not find any significant dependence on the environment in the filamentary and diffuse accretion in our models. This is probably related to a relatively small difference between the overdensities involved, between $\delta\sim 1.3$ and $\delta\sim 3$ (Table\,\ref{tab:DMsim}).

\end{itemize}

In the cold accretion model framework, the filamentary accreting gas can be kept at low temperature within the DM haloes, without being shocked around the virial radius, in constrast with the diffuse accretion in more massive haloes. We have confirmed that the streamer gas temperature remains low, $\sim 10^4$\,K, as it crosses the virial radius, which is substantially lower than the halo virial temperature of $\sim 4-10\times 10^6$\,K, depending on $z_{\rm f}$. Inside the virial radius, some of the gas temperature starts to increase, reaching $\sim 10^{5.5-6}$\,K. The accreting gas is heated up by the galactic winds, is compressed and heated by the decay of the turbulent motions and small-scale shocks. 
However, the amount of the hot gas is not large, the majority of the filamentary flow remains cold, as can be seen in the Figure\,\ref{fig:allfilaT}. Only at $z_{\rm f}=2$, the cold stream appears to be disrupted inside $R_{\rm vir}$, and the filamentary gas is distributed evenly in the range of $10^4-10^6$\,K. 

The properties of diffuse gas differ substantially from that of the filamentary one. Outside $R_{\rm vir}$, most of the gas remains around $10^4$\,K, but inside the halo, the gas heats up and its mass is distributed between $10^5$\,K and up to the virial temperature. 

Figures\,\ref{fig:allfilaflux} and \ref{fig:alldiffuseflux} reveal that about equal mass accretion rates of filamentary and diffuse accretion cross the virial radii. But deeper in the haloes, the diffuse accretion rates start to dominate, as the filamentary flow dissolves gradually. Therefore, the hot gas starts to dominate in mass in the inner halo. 

\citet{field20} have compared in detail the CGM properties in various numerical simulations at $z=0$, invoking the pressure-entropy plane. Following their work, we have attempted to characterize the properties of the gas filling up the CGM in our high resolution simulations into the pressure-entropy plane, separating the inner and outer CGM of three ZLCW haloes (i.e., Z2LCW, Z4LCW and Z6LCW), at $z_{\rm f} = 6,$ 4 and 2 (Figure\,\ref{fig:pressure}). This gas excludes the ISM of the central galaxy determined by the HOP. Moreover, we limit our analysis to CW models residing in the low overdensity regions. 

We find that both CGM regions differ substantially in their properties. Generally, we distinguish three different regions which account for most of the gas mass in Figure\,\ref{fig:pressure}. First, the top region, which extends from $K/K_{\rm vir}\sim 10^1$ to $\sim 10^{-2}$. Second, the middle region of an isothermal gas at $\sim 10^4$\,K at $K/K_{\rm vir}\sim 10^{-2}-10^{-4.5}$. And third, the horizontally extended and, therefore, radially extended, low-entropy gas across a range in density at $K/K_{\rm vir}\sim 10^{-4.5}-10^{-6}$.  

We can reconstruct the locations and masses of this gas in this halo. The total mass of the CGM is $\sim 10^{10}\,M_\odot$ at $z_{\rm f} = 6$, $\sim 3\times 10^9\,M_\odot$ at $z_{\rm f} = 4$, and $\sim 2\times 10^9\,M_\odot$ at $z_{\rm f} = 2$.  The first region defined above contains $\sim 3\times 10^9\,M_\odot$ (inner CGM) and  $\sim 4\times 10^8\,M_\odot$ (outer CGM) of the virialized hot diffuse gas at $z_{\rm f} = 6$. This amount decreases by a factor of 3 (inner CGM) and stays the same (outer CGM) at $z_{\rm f} = 4$. It stays unchanged further on at $z_{\rm f} = 2$.

The second region has $\sim 4\times 10^9\,M_\odot$ (inner CGM) and  $\sim 2\times 10^9\,M_\odot$ (outer CGM) of the nearly isothermal gas at $z_{\rm f} = 6$.  This gas is in thermal equilibrium with the cosmological UV background.  Its amount decreases by a factor of 4 (inner CGM) and stays the same (outer CGM) at $z_{\rm f} = 4$. It further decreases by a factor of 2 (inner CGM)  and by a factor of 10 (outer CGM) at $z_{\rm f} = 2$.

The third region includes the starforming gas locked in the substructures.  Its left boundary is limited by our critical density for star formation, $4\,{\rm cm^{-3}}$.  It extends horizontally to high pressure, and this extension depends on the star formation recipe --- it can be seen in the log\,$T-{\rm log}\,\rho$ diagrams as well. Note also that because this gas is located in substructures, it should be normalized by the virial pressure of each object, but we ignore this change in normalization.   The starforming gas mass has $\sim 5\times 10^8\,M_\odot$ (inner CGM) and  $\sim 4\times 10^7\,M_\odot$ (outer CGM) at $z_{\rm f} = 6$. This amount decreases by a factor of 10 (inner CGM) and by a factor of 100 (outer CGM) at $z_{\rm f} = 4$. It further decreases by a factor of 100 (inner CGM)  and by another factor of 10 (outer CGM) at $z_{\rm f} = 2$. 

When the starforming gas exists,  it has the highest pressure of all three components discussed above. The high entropy gas which is always present in simulations extends from low to intermediate pressure, while the isothermal gas exhibits the lowest pressure and intermediate entropy.

Overall, we observe similar structural details in pressure-entropy plane, as well as substantial differences between the gas properties in the inner and outer CGM, as well as redshift evolution in similar DM haloes. In the inner CGM, the amount of the starforming gas decreases abruptly with $z_{\rm f}$, and this gas is basically absent at $z_{\rm f} = 2$. The amount of $10^4$\,K gas decreases as well with decreasing redshift, but not so dramatically. In the outer CGM, the starforming gas disappears already by $z_{\rm f} = 4$. And the amount of $10^4$\,K gas decreases sharply as well, becoming negligible at $z_{\rm f} = 2$. In summary, the CGM analysis exposes its multiphase character and being away from thermal and dynamic equilibrium.

Note that masses quoted above represent the CGM in specific and similar haloes used in our simulations. The median properties of the CGM, such as mass fractions, have been found to differ among numerical simulations due to the galactic feedback \citep[e.g.,][]{davies20}. Such differences in the CGM become explicit in our nearly identical haloes at three different redshifts due to the changing galactic feedback.

Our results presented in sections\,\ref{sec:results421} and \ref{sec:results422} show that filamentary and diffuse accretion rates appear similar in CW and VW models outside the radius of $\sim 50h^{-1}$kpc.  This appears puzzling as the VW feedback is about 8 times stronger than the CW one (e.g., section\,\ref{sec:winds4}).  However, one must remember that the galactic wind feedback is supplemented by the SN feedback which is strongly related to the star formation rate. 

In order to demonstrate the relative effects of the wind and SN feedback, we have calculated the instantaneous energies of the wind and the SN energy depositions at various times. For example, at $z_{\rm f} = 2$, for the galaxies Z2LCW and Z2LVW, the ratio of the kinetic energies of the VW to CW wind feedback is $\sim 14$. Hence, the wind energy VW is substantially higher than in the CW at this time. Next, we have estimated the energy contribution from the SN during the preceding 1\,Gyr in each model, and found than the ratio for VW to CW energies is $\sim 0.38$, much smaller for the VW model. This is understandable, as the efficiency of the SF in VW models is substantially lower than in CW, resulting in a much larger fraction of gas in galaxies \citep{bi22a}.  Combining the contribution of the wind and the SN rate in both models, we find that the total energy feedback in VW model is only larger by a factor of $\sim 2.6$. Hence, we conclude that the absence of a real difference between the mass accretion rates in VW and CW models is due to the relatively small net difference in the applied feedback.

One should be careful not to make far reaching statements about the state of the accreting gas across the virial radius based on numerical simulations. For example, the AGN feedback, which is absent in the current simulations, has been shown to have a strong effect on the state of the halo gas and hence its accretion rate onto the central galaxy \citep[e.g.,][]{woods14,ubler14,nelson15}. We, therefore, limit our conclusions about the state of the CGM which follow from the types of feedback used here. 

While the inflow velocity profiles along the streamers increase at smaller radii, within $\sim 100h^{-1}$kpc the trend is reversed. The maxima of this velocity decrease with lower $z_{\rm f}$ (Figure\,\ref{fig:allfilavr}). This decrease appears to correlate with the halo escape speed which decreases with time because the halo concentration of similar haloes at different redshifts decreases with lower $z_{\rm f}$. The velocity profiles of diffuse accreting gas are flatter  (Figure\,\ref{fig:alldiffusevr}), but generally follow the same trend as the filamentary accretion.

Slicing the streamers perpendicular to their spines (e.g., Figure\,\ref{fig:filacross}) allows us to define the high density core and low density envelope region(s) in each slice, e.g., Figures\,\ref{fig:fila1edgeon} and \ref{fig:filacross}. The high and low density regions in the streamers display velocity gradients, i.e., the core region separates higher from lower infall velocities. Such gradients are expected to contribute to the ablation process of the core region in streamers by the Kelvin-Helmholtz instability. Indeed, at smaller radii, we observe the splitting in the streamer's core, and the streamer starts to resemble a spaghetti-type flow.  

Along with the gas temperature and density radial profiles, we do find substantial gradients for the metallicity on the halo scale, i.e., between the inner and outer CGM. But on smaller scales, larger variations of metallicity prevail, forming pockets of high metallicty. When a (sub)structure is embedded in a streamer or in diffuse accretion, its metallicity diffuses out either by stellar feedback or other processes, forming these pockets.

As one of the important issues in understanding the evolution of the filamentary inflow inside the virial radius, we have measured the ratio of the cold-to-hot gas mass, $f_{\rm ch} = M_{\rm cold}/M_{\rm hot}$, at $R_{\rm vir}$, $0.5R_{\rm vir}$ and $0.1R_{\rm vir}$ (not shown here). We define the cold gas with $T\ltorder 3\times 10^4$\,K, and hot gas with the temperature above this threshold. We find that $f_{\rm ch}$ declines monotonically with lower $z_{\rm f}$ from $\sim 2-3$ at $z_{\rm f}= 6$ to $\sim 0.25-0.5$ at $z_{\rm f} = 2$ by an order of magnitude. The sharper decline happens inside $0.5R_{\rm vir}$. Not  accidentally this decline coincides with the region where the galactic outflows are strong, i.e., inside $100h^{-1}$kpc. At all radii, $f_{\rm ch}$ falls below unity at $z_{\rm f} = 2$, so, a sharp decrease at lower redshift haloes. Note that our haloes have the same mass by design at all these final redshifts. 

Finally, we analyse the structure of the CGM which populates the entire DM haloes outside the central galaxies. As an example, we present the evolution of the CGM around the Z6LCW galaxy --- its  mass growth, and temperature and metallicity evolution (Figure\,\ref{fig:awesome_image3}). We have divided the CGM which occupies the parent DM halo, into inner and outer CGM. Based on the ratio of cold-to-hot gas, $f_{\rm ch}$, discussed above, the inner CGM can be defined inside the inner $100h^{-1}$kpc, and the outer CGM region between this radius and the virial radius. The inner CGM is also affected by the galactic feedback stronger than the outer CGM. For models presented here, we find that the CGM is not in equilibrium, nor kinematically, thermodynamically, chemically or temporally. It is constantly perturbed by the influx mass, momentum and energy across $R_{\rm vir}$ and from the central galaxy, and forms a multi-phase structure. This conclusion agrees with observations of low-redshift galaxies \citep[e.g.,][and refs. therein]{tomlin17}.

Furthermore, the filamentary flow which does not impact the central galaxy expands to the baryonic backsplash radius, which appears to lie at $\sim 1-2R_{\rm vir}$, smaller than the DM backsplash radius, and, therefore, can provide additional perturbations to the CGM, especially for $z\gtorder 1$, when the filamentary accretion is more significant (e.g., Figure\,\ref{fig:fila1edgeon}, lower panel). The shock associated with the splash radius is visible in our analysis but is not shown here. This result agrees well with previous determinations of the shock position \citep[e.g.,][]{nelson16}.

Recent work has raised an interesting issue that the results of zoom-in simulations can be affected to some extent by the `butterfly effect'\footnote{The butterfly effect has been introduced in relation to a system whose evolution is highly sensitive to the initial conditions, e.g., a chaotic system \citep{bradbury52,lorenz63}.} by introducing stochasticity based on the usage of random number generators implemented in various recipes, such as star formation routines, etc. \citep[e.g.,][]{genel19,ramesh23}. To what extent this affects our results?

We have encountered this effect as well as the `timing effect' \citep{spri08} in another set of the zoom-in numerical simulations (Goddard et al., in prep.). We found that the evolution initially separates, but then runs in parallel, with various events happening with a small time delay. However, we did not find that the evolution diverges as in chaotic systems.  Hence, in the current set of simulations, even if both effects are present, they have only a small amplitude and do not grow with time. For example, the actual differences between the CW and VW models appear not to be affected by them.

\subsection{Streamer properties in the innermost halo}
\label{sec:innerHalo}

Turning our attention to the innermost haloes, inside $\sim 30\,h^{-1}$\,kpc, we observe that the geometrical width of the streamers exceeds substantially the central galaxy cross sections. In all our models, the streamers are inclined to the equatorial plane of the central galaxy. The inclination angle is never around $90^\circ$ or $0^\circ$. For example, Figure\,\ref{fig:fila1join2}, shows the inclination angle being $\sim 30^\circ$ for the streamer \#1 (see also Figure\,\ref{fig:allfil} for the identification of this filament).

Inside this region, the streamers have been observed to dissolve gradually by the ablation process (e.g., Figure\,\ref{fig:fila1join2}). This process is characterized by an increase in the turbulent layer associated with the streamer and decreasing radii. In order to quantify the turbulence, we have calculated the associated vorticity (see section\,\ref{sec:results44})  and have pixelized it (Figure\,\ref{fig:fila1join}). It shows that the dissolution is rapid and the turbulence, which is minimal at higher radii, is spreading sideways. A related question is to what extent this turbulent flow in the innermost halo region contributes to the turbulence in the underlying galactic disk \citep[e.g.,][]{choi13}

We can recognize few types of interactions between the baryonic filamentary inflow and the galactic disk. First, some of the inflow impacts the galaxy and has both prograde or retrograde motion with respect to the rotation of the stellar/gaseous galactic disk. We expect, that the prograde encounter between the filamentary inflow and the galactic disk is smoother than the retrograde one.  Second, and more generally, most of the gas inflow misses the direct encounter with the disk and creates a `tail' flow around the disk, which appears to be turbulent. This part of the inflow which has missed the disk contributes to buildup of the extended and highly turbulent region around the disk, and its kinematics differs substantially from that of the filamentary or diffuse accretion at larger radii.

We observe that all the halos in our simulations are connected typically by 2-3 filaments, and only in one case by 4 filaments. This is in agreement with other simulations (e.g. IllustrisTNG simulation, \cite{nel18}). The reason for this is still not well understood. 

Comparison between our profiles of thermodynamic variables provide a complementary information to those presented in \citet{field20}.  Direct comparison for our median profiles agree with the latter. But our hex-binned distributions of the temperatures in filamentary and diffuse accretion separate their gas mass as a function of temperature. For example, the increasing mass accumulation around $T\sim 10^4$\,K in the inner halo is evident in the filamentary gas (Figure\,\ref{fig:allfilaT}), but not in the diffuse accretion (Figure\,\ref{fig:alldiffuseT}).

The difference in growth rate with redshift in our galaxies (Figure\,\ref{fig:fila2join}) is also mainly due to the time employed by the galaxies to reach their final masses, i.e., for a given mass range, galaxies at high-$z$ grow faster than galaxies at lower-$z$.  

\subsection{From the virial radius to the central galaxy}
\label{sec:CGM}

Finally, we take a look at the big picture of accretion and outflows inside the parent haloes of central galaxies and its immediate environment, i.e., within $\sim 2R_{\rm vir}$. While the number of models presented here is limited compared to the cosmological simulations in large computational boxes, the advantage of our zoom-in simulations is the higher resolution which allows to obtain a detailed picture of kinematic and thermodynamic properties of these flows. 

We find that the gaseous component of DM haloes in our models has a complex structure, both thermodynamically and kinematically. Filamentary and diffuse inflows differ in density, temperature, metallicity, and in the associated velocity field. Moreover, the observed outflows have multiple sources of origin. First come the galactic winds of various strength. Second contribution comes from filamentary accretion which missed the central galaxy and has been diverted radially out. The cross sections of the streamers are typically much wider than those of the galaxies. Much of the flow along the streamers, therefore, misses the galaxy, but turns around and can even interact with itself.  

The outer boundary of this interaction defines the backsplash radius \citep[e.g.,][]{adhikari14}. 
The backsplash radius of the gas is expected to be smaller than the one for the DM due to the shocks and associated dissipation. By using the logarithmic derivative of the density with respect to the radius, we have checked for the position of the backsplash radii for our central galaxies. For $z_{\rm f} = 6$ and 4, we find a good corresponce with $R_{200}$, which lies close to $R_{\rm vir}$. But for $z_{\rm f} = 2$, we find that the backsplash radius lies further out of $R_{\rm vir}$, on the average around $\sim 1.7R_{\rm vir}$. We conclude that the shock associated with the baryonic backsplash radius in our simulations lies at $\sim 1-2R_{\rm vir}$. 

We find that the properties of the filamentary accretion vary with distance to the central galaxy. We detect the ablation process of the filamentary inflow, associated with Kelvin-Helmholtz instability and generation of turbulence around the dissolving filaments, leading to the spaghetti-type flow. Thus a filament separates in a number of individual streamers which continue to be ablated and mixing with the CGM. Hence, we associate the CGM with the entire volume of the DM halo surrounding the central galaxy and define it as the transition region between the galactic ISM and the virial radius. This confirms that the CGM is highly inhomogeneous, i.e., multiphase, in its thermodynamic and kinematic properties.

We have compared our analysis of the filamentary accretion with that performed by \citet{ramsoy21} at high redshifts using the adaptive mesh refinement (AMR) code \textsc{ramses}, with a comparable resolution to ours. Both approaches have detected a developed turbulence measured by vorticity in the filaments. And in both cases, this vorticity has delineated the boundary of a filament, allowing to determine its radius, i.e., its cross section. While Ramsoy et al. have focused on the typical filament structure, we have addressed the dissolution process of the filament. The core-envelope structure of a typical filament in our simulations have been found to separate in a spaghetti-type flow (section\,\ref{sec:results43}). The Kelvin-Helmholtz instability further acted to ablate the streamers, ultimately mixing it with the diffuse environment.

We have also calculated the fraction of volume occupied by the walls shown in Figure\,\ref{fig:allfil}, and found it is roughly independent of the final redshifts, $z_{\rm f}$. By definition, the sheets represent a one-dimensional stable manifold, while a filamentary streamer represents a two-dimensional stable manifold. Note, that \citet{hahn07a} has measured evolution of this fraction and found that it increases untill $z\sim 2$ and decreases thereafter. However, there is no contradiction with our results --- our redshift dependence is not an evolutionary one, but compares the environment of similar mass haloes at different {\it final} redshifts. When we switch to evolutionary redshift for each halo separately, we confirm the Hahn et al. result.    

All the streamers exhibit metallicity in the range of $Z/Z_\odot\sim 10^{-3}-10^{-1}$, with a radial gradient between the outer and inner halo. Inside $R_{\rm vir}$, the streamers have a mass-weighted metallicity $Z/Z_\odot\gtorder {\rm few}\times 10^{-2} - {\rm few}\times 10^{-1}$, and agrees well with observations in the optical-X-ray bands \citep[e.g.,][]{werk14,lehner19}. For the diffuse accretion, the metallicity is smaller by a factor of a few.

\section{Conclusions}
\label{sec:conclusions}

Based on a set of high-resolution zoom-in cosmological simulations, we investigate the baryonic filaments (streamers) and the diffuse accretion flows which channel the gas across the virial radii and down to the galactic regions, under a dual action of accretion and galactic outflows. We choose a set of DM haloes with similar masses of log\,$({M_{\rm vir}}/{\textrm M_\odot}) \sim 11.65$ at three representative redshifts, $z_{\textrm f} = 6$, 4 and 2 from \citet{bi22a}. These haloes have been evolved in relatively low and high overdensities compared to the average density in the universe, and using two different stellar feedback.

Using a hybrid d-web/entropy method supplemented by kinematics in the innermost regions, we have mapped the filamentary streamers and seprated them from galactic outflows and diffuse accretion flows. This allowed us to analyze the dynamic and thermodynamic properties of the CGM in nearly identical haloes at different redshifts. We find that the CGM is highly inhomogeneous and multiphase, and not in thermodynamic or dynamic equilibrium. 

We find that accretion rates in filamentary streamers exhibit decrease with decreasing redshifts, and the inflow velocities along these filaments decrease by a factor of $\sim 2$ with lower $z$, again in similar haloes. The temperature inside the CGM increases at smaller radii, as well as with decreasing redshit. 

The filamentary streamers display a core-envelope structure inside the virial radius --- a higher density, lower temperature core surrounded by a lower density, higher temperature envelope. The filament separates into spaghetti-type flow. Inside the inner $\sim 30\,h^{-1}$kpc, we show that the filaments develop the Kelvin-Helmholtz instability, which triggers turbulence, ablates and dissolve them. 

We find that the galactic outflows in tandem with diverted accretion flow affect the accretion flow, mostly within the inner CGM of $\sim 0.5R_{\rm vir}\sim 100h^{-1}$kpc. Finally, we find that the thermodynamic properties of the CGM gas can be separated into three phases on the pressure-entropy plane (first used by \citet{field20}). These regions include the high entropy, low-density hot gas, isothermal gas at $\sim 10^4$\,K in equilibrium with the UV background, and the low entropy starforming gas.

\section*{Acknowledgements}

We thank Phil Hopkins for providing us with the latest version of the code. We are grateful to Alessandro Lupi for his help with GIZMO, and to Peter Behroozi for clarifications about ROCKSTAR. I.S. acknowledges insightful discussions with Nick Kaiser during his stay at the Kavli Institute for Theoretical Physics (KITP), and is grateful for a generous support from the International Joint Research Promotion Program at Osaka University. I.S. acknowledges the hospitality of KITP where part of this research has been conducted. This work has been partially supported by the JSPS KAKENHI grant 16H02163 to I.S., and by the NSF under Grant No. NSF PHY-1748958 to KITP.    The STScI is operated by the AURA, Inc., under NASA contract NAS5-26555. E.R.D. acknowledges support of the Collaborative Research Center 956, subproject C4; and from the Collaborative Research
Center 1601 (SFB 1601 sub-project C5), both funded by the Deutsche
Forschungsgemeinschaft (DFG, German Research Foundation) – 500700252. Simulations have been performed using generous allocation of computing time on the XSEDE machines under the NSF grant TG-AST190016, and by the University of Kentucky Lipscomb Computing Cluster. We are grateful for help by Vikram Gazula at the Center for Computational Studies of the University of Kentucky.



\section*{Data Availability}

The data used for this paper can be available upon reasonable request. 
 






\bsp	
\label{lastpage}
\end{document}